\documentclass{article}
\usepackage{amsmath}
\usepackage{amsfonts}
\usepackage{graphicx}
\usepackage{hyperref}
\usepackage{comment}
\usepackage{longtable}
\usepackage{caption}
\usepackage[a4paper, total={6in, 9in}]{geometry}

\title{Modelling and Optimization Based Control for Demand Response in Active Distribution Networks}
\author{Arsam Aryandoust\textsuperscript{1, 2}}

\begin{document}

\maketitle
\begin{center}
    \textsuperscript{1}Laboratory for Information \& Decision Systems, MIT, United States. \\
    \textsuperscript{2}Climate Policy Lab, ETH Zurich, Switzerland. \\
    arsam@mit.edu
\end{center}

\begin{abstract}
    We explore how Demand Response (DR) can effectively provide electricity system services such as for the management of bi-directional power flows and the control of voltage deviations in active distribution networks, without compromising consumer comfort or adversely affecting grid operation in the transmission network. By translating intricate power system physics into straightforward control objectives, we design DR control algorithms that can operate within realistic computational time frames at scale. We conduct simulation-based experiments and find that minimizing the Euclidean-Norm of the total residual load at transformer sub-stations is an effective objective for harnessing DR for the dispatch of renewable electricity grids. We show that this control objective can be efficiently pursued in sequential order, without optimal power flow calculations or information about the topology of a grid. Additionally, we find that pursuing this objective reduces the sum of peak power flows along all lines in active distribution networks, and can therefore enable both lower transmission losses and lower voltage deviations.
\end{abstract}

\newpage
\section{Introduction}

We have changed the balance of our planet by burning fossil fuels and emitting large amounts of greenhouse gases. The growing incidents of floods, storms and droughts are early indicators of the extreme weather conditions that we can anticipate more frequently due to an escalating global warming and climate change. The urgency to tackle climate change has motivated national leaders from around the world to unite and devise collective strategies \cite{Paris2015}. One critical and immediate strategy for tackling climate change is to transition the majority of our energy consumption to electricity and supply this electricity mainly from renewable energy sources like wind and solar \cite{Shukla2022}.

The shift from fossil fuels to renewable energy sources presents several operational challenges for our contemporary power systems. Traditional large-scale power plants are gradually giving way to decentralized energy sources like rooftop photovoltaic (PV) and micro-wind units. As a result, distribution grids, designed solely for accommodating the consumption of electricity over the past century, now also have to accommodate the generation of electricity. These are therefore often referred to as \emph{active distribution networks}. The emergence of active distribution networks introduces a set of technical challenges, including the occurrence of bi-directional power flows, the variability in energy supply due to the intermittent nature of wind and solar sources, and a reduction in the stabilizing inertia traditionally provided by the rotating masses of conventional power generators.

Maintaining a stable equilibrium in power systems requires precise matching of consumption and generation, which regulates the frequency in a system, while also ensuring that power flows, voltage deviations and phase imbalances remain within stable operational bounds. Bi-directional power flows amplify the intricacy of this coordination. Fluctuations from wind and solar energy additionally increase the uncertainty and reduce the controllability of power generation, which has previously been used to match the already inherent uncertainty of power consumption. The absence of inertia from rotating masses further reduces a power system's inherent capability to stabilize operations by automatically adjusting the rotational speed of generators within swift response times. In order to ensure a harmonious balance of consumption and generation in future electricity grids that are dominated by renewable energy, we need to increase the amount of electric storage capacities and develop fast computational solutions for dispatching a large number of these capacities in real-time.

An effective way for reducing the demand of new storage capacities is to utilize available Demand Response (DR) potentials. DR incorporates the use of flexible load for providing electricity system services on short time frames \cite{Aryandoust2017}. With the roll-out of Smart Meters and the Internet of Things (IoT), DR resources are increasingly tapped in the domestic, industrial and commercial sectors, and can be operated along with electric storage capacities from for example electric vehicles \cite{Aryandoust2019, Lin2023} and electric batteries installed in buildings \cite{Elio2023}. Given that most of these DR resources are already integrated into our concurrent electricity grids, they present both a sustainable and cost-effective way to reduce the demand for new storage capacities. However, optimizing DR resources at a large scale brings forth a pivotal research question: What is an effective control objective for utilizing a large number of DR resources for the dispatch of active distribution networks, without compromising consumer comfort or adversely affecting gird operation in the remaining transmission network?

Studies that come closest to answering this question primarily focus on DR for voltage control in distribution networks \cite{Petinrin2014, Christakou2014, Zakariazadeh2014, Zhu2019, Sharma2021, Gupta2022}. For instance, Christakou et al. develop a method to determine the optimal settings for transformers and active/reactive power injections at each bus in a network \cite{Christakou2014}. They utilize a thermal model for flexible loads to reduce voltage deviations. However, their algorithm relies on having access to detailed information about active/reactive power injections and phase angles at each bus in the network. Zakariazadeh et al. optimize DR dispatch and its associated costs for corrective voltage control through a multi-objective optimization approach, taking into account optimal power flow \cite{Zakariazadeh2014}. Once again, this method necessitates comprehensive data on the topology and state of the electric grid. Zhu et al. employ a two-step process \cite{Zhu2019}. They first perform optimal power flow calculations for the transmission system and then dispatch DR resources alongside PV outputs, capacitors, and inverters in the distribution system to meet optimal active and reactive power injections. This is done without compromising voltage stability in the distribution system. Similar to the other approaches, their algorithm also requires extensive information about the distribution system's topology and may not necessarily enhance voltage stability margins in active distribution networks. 

It is worth noting that all existing methods involve solving the optimal power flow problem, which can be particularly challenging for large networks with a multitude of controllable DR resources in real-time. Additionally, they often depend on information about the precise topology of the grid, which can be hard to obtain for low-voltage distribution grids that electricity system operators often operate as black boxes. These challenges persist as we are lacking effective control objectives that could alleviate the need for optimal power flow calculations and detailed grid topology information.

Here, we fill this research gap by translating the intricate physical relationships of a power system into simple control objectives that alleviate the need for optimal power flow computation. We yield two objectives that achieve such simplifications. The first objective we yield alleviates the need for information about the topology of a grid and its state at every bus for dispatching DR resources. The second objective we yield attempts to leverage information about the topology of a grid and its state at every bus to enhance control decisions. In the following sections, we outline the flow of information and control scheme that we assume for this purpose. Next, we describe our methodology for implementing five different DR control algorithms, each pursuing a variant of our simplified objectives. We then proceed with presenting and discussing our experimental results.

\subsubsection*{Control scheme}

We assume a bottom-up propagation of information that starts at the device-level on the low-voltage distribution grid and ends at the system-operator-level on the high-voltage transmission grid. On the device-level, we use a flexible load model that is comparable with the thermal load model proposed by Christakou et al. \cite{Christakou2014}. Our model generalizes to non-thermal loads and can therefore represent a wider range of DR applications. This allows us to account for consumer comfort constraints in our simulations, and evaluate DR control algorithms that do not compromise these. We assume that DR resources are first dispatched for voltage control in low-voltage distribution grids. To the extent that resulting voltage stability margins permit, additional DR resources are then assumed to be passed to the respectively higher voltage levels for system operators to control power flows, phase imbalances and eventually also frequency. In order to communicate available DR potentials throughout the different voltage levels from the DR application on a device-level to distribution system operators and eventually to transmission system operators, we can for example assume the use of load shifting potential profiles as proposed by Aryandoust et al. \cite{Aryandoust2017}. An effective way to send control actions to a large number of DR applications in a grid can in contrast be reversed to a top-down approach as proposed by Christakou et al. \cite{Christakou2014}.

The topology of a power system on the high- and medium-voltage level is well-known and its state is well-measured or relatively easy to estimate. The number of buses is further often manageable for optimal power flow calculations in real-time. Here, we therefore focus on controlling bi-directional power flows and voltage deviations on the low-voltage distribution grid. We further neglect the prediction step of residual electric load based on its historic measurements in addition to weather forecasts, as this is an extensively studied field with advanced machine learning solutions \cite{Aryandoust2022}.

\section{Methods}

In this section, we present our methodology for developing and empirically testing our developed DR control algorithms. First, we introduce a model that is able to represent arbitrary DR applications from the domestic, commercial and industrial sectors; we first present the general model and then provide explicit numerical examples for representative applications from the domestic sector. Second, we describe the two simulation scenarios that we use to empirically test our DR control algorithms. Third, we derive our simplified control objectives from complex power flow relationships. Fourth, we outline the formulations of the mathematical optimization problems that allow us to pursue these objectives; we discuss six different properties of these problems and the different choices we can make on each of these. Fifths, we present our complete DR control algorithms, which incorporate solutions to our formulated optimization problems and hence also the pursuit of our simplified control objectives; we propose five algorithms and provide exemplar mathematical formulations for these.

\subsection{Nomenclature}

The following list contains all important symbols that we use throughout this work:

\begin{longtable*}{ll}
    $T$ & number of simulation time steps \\
    $P$ & active power \\
    $Q$ & reactive power \\
    $cos(\phi)$ & load factor \\ \\

    $E_i$ & voltage phasor at bus $i$ \\
    $U_i$ & voltage magnitude at bus $i$ \\
    $\theta_i$ & voltage angle at bus $i$ \\
    $g_{km}$ & real part of admittance between buses k and m \\
    $b_{km}$ & imaginary part of admittance between buses k and m \\ 
    $b_{km}^{sh}$ & imaginary part of admittance of shunt element between buses k and m \\ 
    $P_{km}^{line-loss}$ & active power loss on line between buses k and m \\
    $Q_{km}^{line-loss}$ & reactive power loss on line between buses k an m \\ \\
    
    $T_{set}$ & set point temperature \\
    $T_{db}$ & dead-band temperature \\
    $T_{ss}$ & steady state temperature \\
    $T_{LOL}$ & lower operational temperature limit for heat pumps \\
    $T_{HOL}$ & higher operational temperature limit for heat pumps \\
    $T_{DB}$ & dead-band temperature total range \\
    $T_{up}$ & upper storage temperature bound for consumer comfort \\
    $T_{low}$ & lower storage temperature bound for consumer comfort \\ \\
    
    $c_{use}$ & constant thermal usage parameter \\
    $c_{sol}$ & constant solar gain parameter \\
    $c_{los}$ & constant thermal loss parameter \\
    $c_{inp}$ & constant electric input parameter \\ \\

    $x_t$ & internal storage state \\
    $x_t^{inp}$ & input/charging of storage via electric consumption \\
    $x_t^{out}$ & output/usage of storage  \\
    $x_t^{loss}$ & loss such as through insulation \\ \\ 
    
    $s_t^{tem}$ & time series for ambient temperature forecasts \\
    $s_t^{sol}$ & time series for solar gain forecasts \\
    $s_t^{wat}$ & time series for hot water demand forecasts \\
    $s_t^{occ}$ & time series for occupancy of household forecasts \\ \\
    
    $u_t$ & power switch thermal device (On/Off) \\
    $v_t$ & internal controller switch (On/Off) \\
    $a_t$ & relaxation factor for consumer comfort \\ \\
    
    $r_t^{act}$ & non-controllable active residual load \\
    $r_t^{react}$ & non-controllable reactive residual load \\
    
\end{longtable*}

\subsection{Demand Response application model}
We can represent arbitrary DR applications as both open and closed thermodynamic systems. Each application has a maximum storage capacity that depends on the size and material of its internal storage matter. Regardless of application-dependent specifications of internal storage, we can define the storage state $x_{t+1}$ at any time instance $t+1$ in dependence on the previous storage state $x_t$, storage usage/output $x_t^{out}$, losses $x_t^{loss}$ and charging/input via electric consumption $x_t^{inp}$ as 

\begin{equation*} 
    x_{t+1} = x_t - x_t^{out} - x_t^{loss} + x_t^{inp}.
\end{equation*}

\begin{figure} [ht]
    \centering
    \includegraphics[height=6cm]{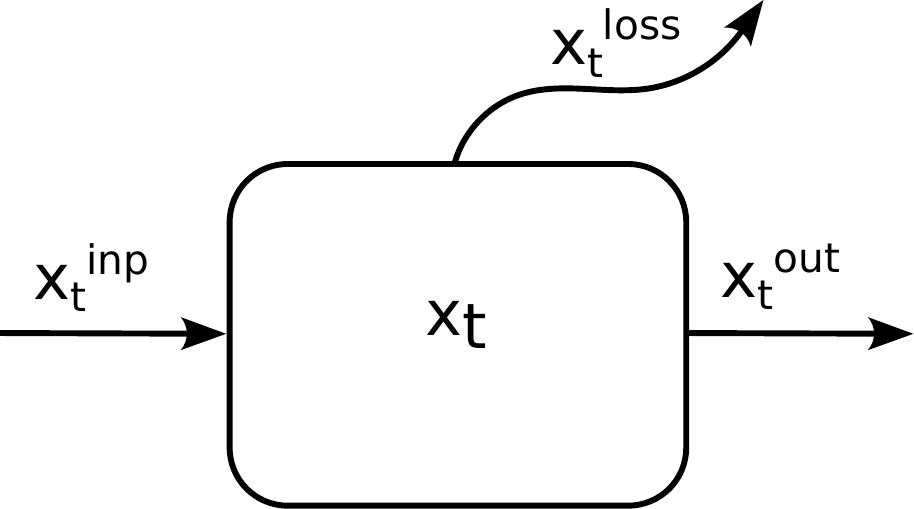}
    \caption{Storage model for arbitrary Demand Response applications. $x_t$ represents storage state. $x_t^{inp}$ represents storage input via electric consumption. $x_t^{out}$ represents storage output through usage. $x_t^{loss}$ represents losses such as through insulation of thermal storage matter.}
    \label{fig:DR_application}
\end{figure}

This allows us to express arbitrary behaviours for inputs, outputs and losses of a storage system using both hand-designed and learned functional relationships. We can generally express these states of charge in percentage (\%), and represent arbitrary flexible loads that can be used for DR. Figure \ref{fig:DR_application} visualizes the respective model. In the following sub-sub-sections, we introduce a number of DR applications from the domestic sector that we use for our simulations. For each DR application, we define a concrete functional relationships used for inputs, outputs and losses. We further provide the concrete numerical values we use for our experiments. All DR applications we use here contain thermal storage capacities. The DR model we propose, however, is able to realistically represent a wide range of alternative applications from the industrial and commercial sectors as well \cite{Aryandoust2017}. 

We represent constant parameters by $c_{i}$ with $i \in [1, 2, 3, 4]$ and time series of predicted values by $s_{t}^{(j)}$ with $(j) \in [tem, sol, wat, occ]$. Various application-dependent temperatures are described by $T_{(k)}$ where $(k) \in [set, db, ss, LOL, HOL, DB, up, low]$. Furthermore, $cos(\phi)$ describes the load factor, $P$ the active power rating and $Q$ the reactive power rating of an application. Without loss of generality, we use temperature in °C as the internal state of charge in all applications, as these represent thermal loads in all of our examples. Constant parameters contain the respective conversion units.

We use hand-designed parameters for each application that we tune in two steps: first, we scale the output and loss of the storage unit in a way that a full discharge covers the demand of a realistic time period, for example, an entire day for heaters, at the hardest usage conditions, for example, during the coldest winter day given the highest occupancy; second, we scale the input in a way that the maximum occurring output can be met within the same time step for the given power rating of the heater. In reality, however, devices do not always meet their user expectations perfectly. Hence, we create a variety of different over- and under-dimensioned devices by assigning randomly distributed model parameters around the subsequently determined values within a range of +/- 10\% to each device. The chosen model parameters are therefore highly dependent on the time series we use for ambient temperature $s_t^{tem}$, solar irradiance $s_t^{sol}$, hot water demand $s_t^{wat}$, and occupancy $s_t^{occ}$.

\subsubsection{Night storage heater} 
Night storage heaters mostly use a ceramic core for internal heat storage which reaches temperatures of around 600 $^{\circ}$C. The insulation loss is assumed to increase proportional to the deviation between internal state temperature $x_t$ and the constant steady state temperature $T_{ss}$. The thermal usage is given whenever the devices heat a building. Hence $x_t^{out}$ is dependent on ambient temperature $s_t^{tem}$ and heat gain through solar irradiance $s_t^{sol}$. Heat is injected whenever the internal controller switch v\textsubscript{t} and the power switch u\textsubscript{t} are simultaneously turned on and both assume a value of 1. We can write our model equations as:

\begin{align*}
    x_t^{out} &= max(0, c_{use}   (T_{ss} - s_t^{tem}) - c_{sol}   s_t^{sol}) \\
    x_t^{loss} &= c_{los}   (x_t - T_{ss}) \\
    x_t^{inp} &= c_{inp}   P   u_t   v_t
\end{align*}

Our model parameters are further set to:

\begin{align*}
    cos(\phi) &= 0.9 \\
    P &= 5 kW \\
    Q &= P \sqrt{1/cos(\phi)^2-1} = 2.4 kVAr \\
    T_{set} &= 580 ^{\circ}C \\
    T_{db} &= 20 ^{\circ}C \\
    c_{sol} &= 0.001 \frac{^{\circ}C}{W/m^2} \\
    c_{los} &= 0.01 \\
    c_{inp} &= 1.2916 \frac{^{\circ}C}{kW}
\end{align*}

This leads to input, output and loss behaviours of:

\begin{align*}
    x_t^{out} &= -600^{\circ}C/day \\
    x_t^{loss} &= -5.8 ^{\circ}C/h &&\text{with } x_t = 600^{\circ}C \\
    x_t^{inp} &= 6.458 ^{\circ}C/15min &&\text{with } P = 5 kW
\end{align*}

\subsubsection{Storage water boiler}
Storage water boilers use water as storage matter with temperatures between 50 - 70 $^{\circ}$C. Depending on the building and consumers they serve, storage capacities vary significantly and are mostly between 50 and several thousand liters for residential consumers. The thermal usage is strongly dependent on patterns of hot water demand $s_t^{wat}$ of the consumers they supply. We can write our model equations as:

\begin{align*}
    x_t^{out} &= c_{use}   s_t^{wat} \\
    x_t^{loss} &= c_{los}   (x_t - T_{ss}) \\
    x_t^{inp} &= c_{inp}   P   u_t   v_t 
\end{align*}

We set our model parameters to:

\begin{align*}
    cos(\phi) &= 0.9 \\
    P &= 4 kW \\
    Q &= P \sqrt{1/cos(\phi)^2}-1 = 1.3 kVAr \\
    T_{set} &= 70 ^{\circ}C \\
    T_{db} &= 25 ^{\circ}C \\
    c_{use} &= 0.0142 \frac{^{\circ}C}{l} \\ 
    c_{los} &= 0.0125 \\
    c_{inp} &= 1.26 \frac{^{\circ}C}{kW}
\end{align*}

This leads to input, output and loss behaviours of:

\begin{align*}
    x_t^{out} &= -95^{\circ}C/day \\
    x_t^{loss} &= -3.75 ^{\circ}C/h && \text{with } x_t = 95^{\circ}C \\
    x_t^{inp} &= 5.03 ^{\circ}C/15min && \text{with } P = 4 kW
\end{align*}

\subsubsection{Heat pump}
\begin{figure} [ht] 
    \centering
    \includegraphics[height=7cm]{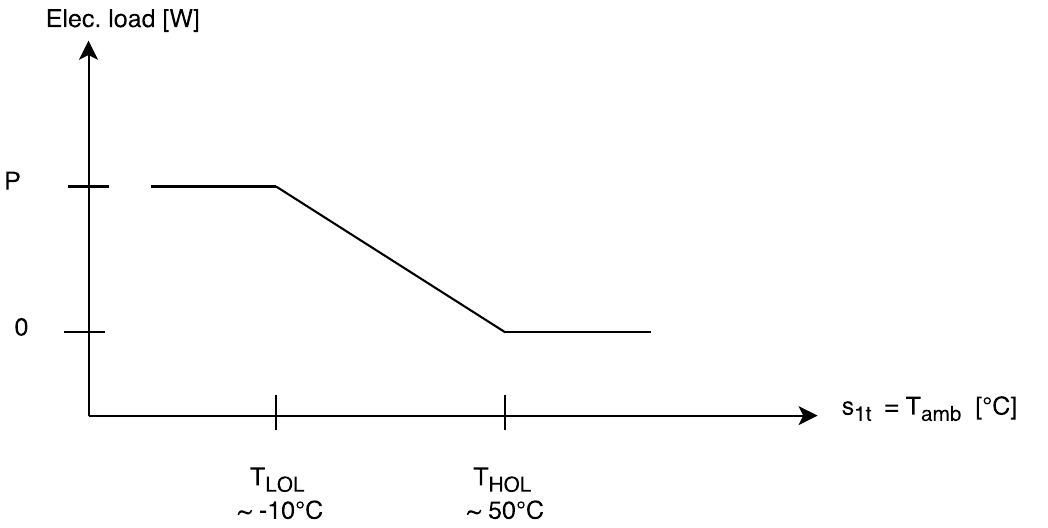}
    \caption{Heat pump input behavior}
    \label{fig:heat_pump}
\end{figure}

Heat pumps can draw thermal energy from different sources and are distinguished as air-source, ductless mini-split, absorption  and geothermal heat pumps. Depending on the type, they can be used for both hot water preparation and room heating. Here, we consider a heat pump with a storage unit for hot water which prepares water on temperatures of around 50 $^{\circ}$C and covers both room heating and hot water demand. The consumption changes proportional in the range of a higher operational temperature limit $T_{HOL}$ and a lower operational temperature limit $T_{LOL}$, and stays constant for all ambient temperatures outside of this range (Figure \ref{fig:heat_pump}). We can write our model equations as:

\begin{align*}
    x_t^{out} &= max(0, c_{11}   (T_{ss} - s_t^{tem}) - c_{sol}   s_t^{sol}) + c_{12}   s_t^{wat} \\
    x_t^{loss} &= c_{los}   (x_t - T_{ss}) \\
    x_t^{inp} &= c_{inp}   P   \frac{(1 - max(0, min( s_{21} - T_{LOL}, T_{HOL} - T_{LOL}))}{(T_{HOL} - T_{LOL}))}   u_t   v_t
\end{align*}

We set our model parameters to:

\begin{align*}
    cos(\phi) &= 0.8 \\
    P &= 5 kW \\
    Q &= P  \sqrt{1/cos(\phi)^2}-1 = 3.1kVAr \\
    T_{set} &= 45 ^{\circ}C \\
    T_{db} &= 5 ^{\circ}C \\
    c_{11} &= 0.01 \\
    c_{12} &= 0.006  \frac{^{\circ}C}{l} \\
    c_{sol} &= 0.001  \frac{^{\circ}C}{W/m^2} \\
    c_{los} &= 0.001 \\
    c_{inp} &= 0.456
\end{align*}

This leads to input, output and loss behaviours of:

\begin{align*}
    x_t^{out} + x_t^{loss} &= -50^{\circ}C/day \\
    x_t^{inp} &= 2.0259 ^{\circ}C/15min &&\text{with } P = 5 kW \\
\end{align*}

\subsubsection{Fridge and Freezer}
Fridges and Freezers use refrigerants and the internal items as thermal matter. Fridges are usually operated at temperatures around 7 $^{\circ}$C and freezers around -18 $^{\circ}$C. We write our input, output and loss equations for both appliances as:

\begin{align*}
    x_t^{out} &= - c_{use}   s_t^{occ} \\
    x_t^{loss} &= - c_{los}   (T_{ss} - x_t) \\
    x_t^{inp} &= - c_{inp}   P   u_t   v_t
\end{align*}

For fridges, we set our parameters to:

\begin{align*}
    cos(\phi) &= 0.7 \\
    P &= 0.8 kW \\
    Q &= P \sqrt{1/cos(\phi)^2-1} = 0.82 kVAr \\
    T_{set} &= 8.5 ^{\circ}C \\
    T_{db} &= 1.5 ^{\circ}C \\
    c_{use} &= 0.0015 ^{\circ}C \\
    c_{los} &= 0.003 \\
    c_{inp} &= 0.3201 \frac{^{\circ}C}{kW}
\end{align*}

Our input, output and loss behaviours assume:

\begin{align*}
    x_t^{out} + x_t^{loss} &= 12^{\circ}C/day \\
    x_t^{inp} &= -0.2561 ^{\circ}C/15min &&\text{with } P = 0.8 kW
\end{align*}

For freezers, we set parameters to:

\begin{align*}
    cos(\phi) &= 0.7 \\
    P &= 1 kW \\
    Q &= P   \sqrt{1/cos(\phi)^2-1} = 1.02 kVAr \\
    T_{set} &= -16.5 ^{\circ}C \\
    T_{db} &= 1.5 ^{\circ}C \\
    c_{use} &= 0.001 ^{\circ}C \\
    c_{los} &= 0.004 \\
    c_{inp} &= 0.2967 \frac{^{\circ}C}{kW}
\end{align*}

Our equations then assume:

\begin{align*}
    x_t^{out} + x_t^{loss} &= 20^{\circ}C/day \\
    x_t^{inp} &= -0.2967 ^{\circ}C/15min &&\text{with } P = 1 kW
\end{align*}

\subsubsection{Air conditioner} 
Air conditioners use the building and internal items and air as thermal storage matter. Temperatures close to 20 $^{\circ}$C are sensed as comfortable and as losses can be very high, only short time periods are suitable for DR operations. We write model equations as:

\begin{align*}
    x_t^{out} &= - c_{use}  s_t^{occ} \\
    x_t^{loss} &= - c_{los}  (s_t^{tem} - x_t) - c_{sol}  s_t^{sol} \\
    x_t^{inp} &= - c_{inp}  P  u_t  v_t
\end{align*}

We set model parameters to:

\begin{align*}
    cos(\phi) &= 0.6 \\
    P &= 2 kW \\
    Q &= P \sqrt{1/cos(\phi)^2-1} = 2.67 kVAr \\
    T_{set} &= 21 ^{\circ}C \\
    T_{db} &= 1.5 ^{\circ}C \\
    c_{use} &= 0.0017 ^{\circ}C \\
    c_{sol} &= 0.00016 \\
    c_{los} &= 0.02 \\
    c_{inp} &= 0.2829 \frac{^{\circ}C}{kW}
\end{align*}

With these, our equations become:

\begin{align*}
    x_t^{out} + x_t^{loss} &= 12^{\circ}C/day \\
    x_t^{inp} &= -0.5659 ^{\circ}C/15min &&\text{with } P = 2 kW
\end{align*}

\subsection{Simulation scenarios}

\begin{figure} [!ht] 
    \centering
    \includegraphics[height=13cm]{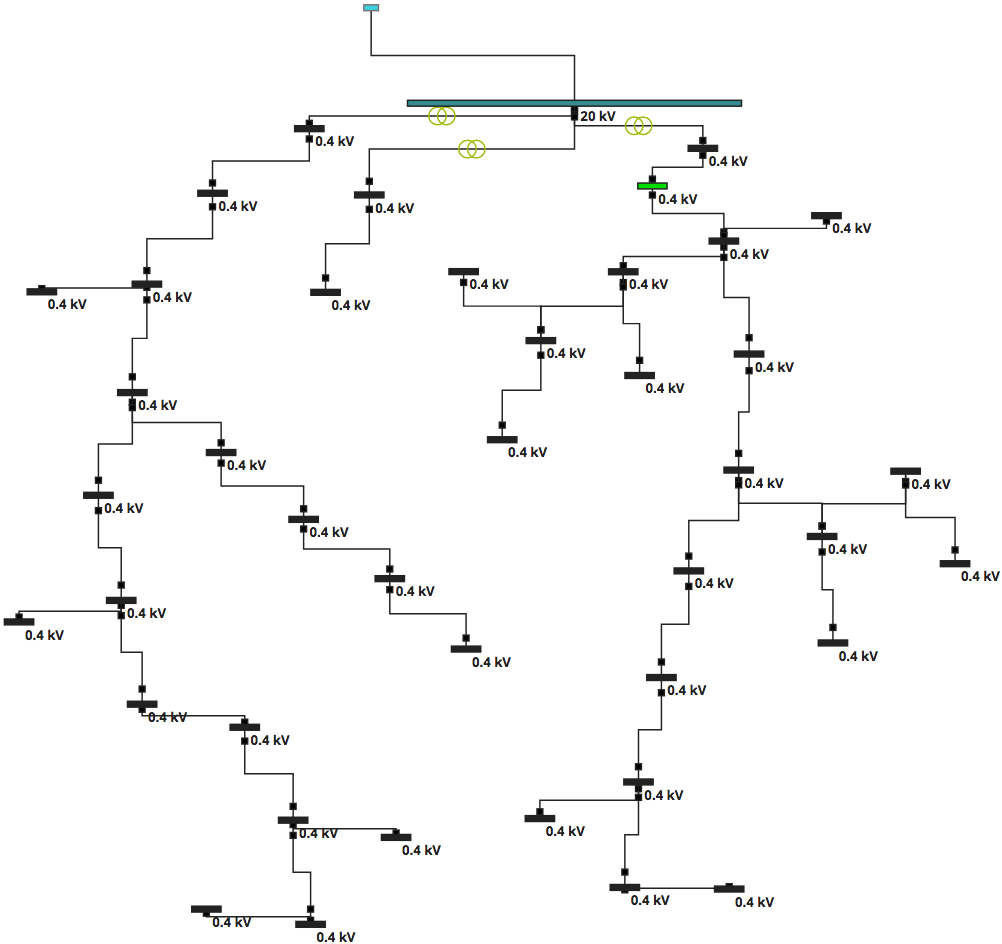}
    \caption{41-Bus low-voltage, radial distribution network that we use for our simulations.}
    \label{fig:grid}
\end{figure}

We implement all simulations with the \emph{Adaptricity.sim} software, previously named \emph{DPG.sim} (\url{https://www.adaptricity.com}). We consider a 41-bus radial distribution network which consists of three $0.4 kV$ low-voltage grids that are connected to the medium-voltage grid through three off-load tap-changing transformers. Figure \ref{fig:grid} visualizes the structure of the grid we use. Tables \ref{tab:grid_data} and \ref{tab:line_data} further provide the specifications of each transformer and line in our simulations. 

\begin{figure}[!ht]
    \centering
    \includegraphics[height=11cm]{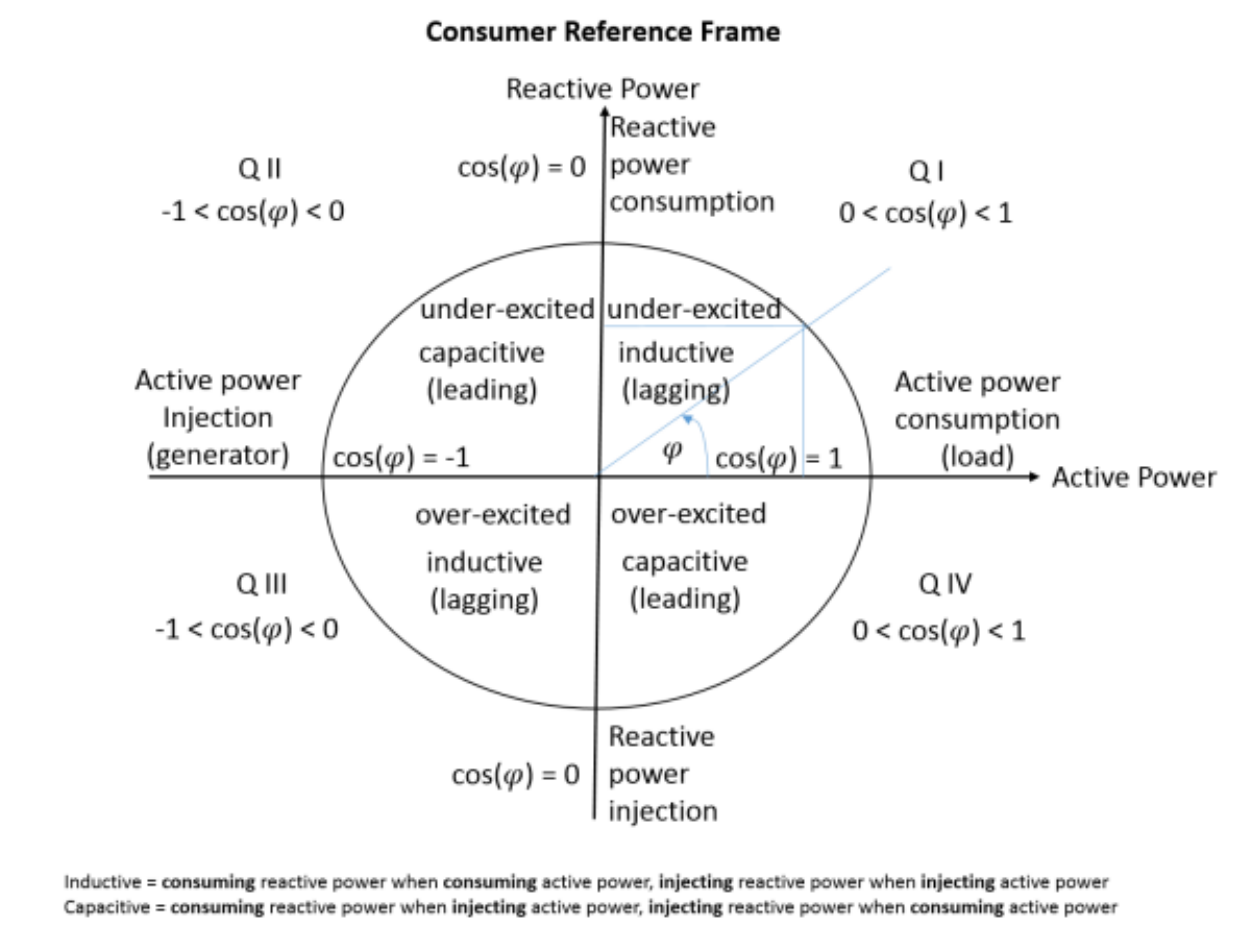}
    \caption{Load factor for active and reactive power (source: \url{https://www.adaptricity.com/}).}
    \label{fig:load_factor}
\end{figure}

We randomly distribute a number of non-controllable loads, PV and micro-wind generation units with fixed consumption and generation profiles on all buses. Together, they represent the non-controllable active and reactive residual load. Load and generation units are randomly given an inductive load factor between 0.8 and 1. All under-excited units like controllable and non-controllable loads consume active and reactive power, while all over-excited units like PV and wind generation units generate active and reactive power. Figure \ref{fig:load_factor} provides an overview for these relationships. We scale the residual load such that active and reactive power fluctuate between both positive and negative values. Our DR resources are represented by a number of the above modelled applications that we randomly distribute on the grid. In absence of external control, the load patterns of our DR resources are regulated by their internal device controllers. Non-controllable residual load and the controllable load from DR resources together represent the total active and reactive residual load.

We simulate two scenarios. First, a randomized scenario, and second, a smoother realistic scenario. In the randomized scenario, our goal is to present a high degree of bi-directional power flows so as to challenge the DR control algorithms that we want to test. This allows us to examine the robustness and generality of our developed DR control algorithms. In the regular scenario, we investigate a future electricity grid with high shares of PV and wind generation. In the following, we describe these two scenarios in further detail. 

\begin{center}
    \begin{table}[!ht]
        \caption{Specifications of the three transformers we use in our simulations.}
        
        \begin{tabular}{| l | l  l l |}
        \hline
        Trafo ID & 218979 & 218645 & 218941\\ 
        \hline
        Voltage 1 (kV) & 20 & 20 & 20\\ \cline{1-1}
        Voltage 2 (kV) & 0.4 & 0.4 & 0.4\\ \cline{1-1}
        Rated power (MVA) & 0.55 & 0.5 & 0.5\\ \cline{1-1}
        Maximum apparent power (MVA) & 0.55 & 0.5 & 0.5\\ \cline{1-1}
        Short Circuit Voltage $U_k (\%)$ & 4.09 & 4.13 & 4.1\\ \cline{1-1}
        Copper losses $U_r (\%)$ & 0.993 & 1.002 & 1\\ \cline{1-1}
        Turns ratio primary side & 1 & 1 & 1\\ \cline{1-1}
        Turns ratio secondary side & 1 & 1 & 1\\ \cline{1-1}
        \hline
        \end{tabular}
        
    \label{tab:grid_data}
    \end{table}
\end{center}

\begin{center}
    \begin{table} [!ht]
        \caption{Specifications of all lines in the grid that we use in our simulations.}
        
        \begin{tabular}{| l | l | l | l | l |}
        \hline
        Line ID & Resistance($\Omega/km$) & Reactance($\Omega/km$) & Length($km$) & $I_{max}(A)$ \\
        \hline
        219009  & 0.397 & 0.279 & 0.03 & 199 \\
        219014  & 0.397 & 0.279 & 0.03 & 199 \\
        219038  & 0.574 & 0.294 & 0.03 & 140 \\
        218765  & 0.284 & 0.083 & 0.035 & 318 \\
        219043  & 0.574 & 0.294 & 0.03 & 140 \\
        219048  & 0.41 & 0.071 & 0.03 & 240 \\
        219053  & 0.41 & 0.071 & 0.03 & 240 \\
        218770  & 0.284 & 0.083 & 0.035 & 318 \\
        218775  & 0.284 & 0.083 & 0.035 & 318 \\
        218780  & 0.497 & 0.086 & 0.035 & 193 \\
        219331  & 0.41 & 0.071 & 0.03 & 240 \\
        219063  & 0.41 & 0.071 & 0.03 & 240 \\
        218785  & 0.497 & 0.086 & 0.035 & 193 \\
        218790  & 0.497 & 0.086 & 0.035 & 240 \\
        219348  & 0.41 & 0.071 & 0.03 & 199 \\
        218804  & 0.822 & 0.077 & 0.03 & 199 \\
        219108  & 0.397 & 0.279 & 0.03 & 199 \\
        219113  & 0.397 & 0.279 & 0.03 & 101 \\
        219118  & 1.218 & 0.318 & 0.03 & 49 \\
        218832  & 3.69 & 0.094 & 0.03 & 240 \\
        821053  & 0.41 & 0.071 & 0.03 & 134 \\
        821045  & 0.871 & 0.081 & 0.03 & 101 \\
        219123  & 1.218 & 0.318 & 0.03 & 318 \\
        218864  & 0.284 & 0.083 & 0.035 & 318 \\
        218869  & 0.284 & 0.083 & 0.035 & 318 \\
        218874  & 0.284 & 0.083 & 0.035 & 318 \\
        218879  & 0.284 & 0.083 & 0.035 & 240 \\
        219162  & 0.41 & 0.071 & 0.03 & 240 \\
        219167  & 0.41 & 0.071 & 0.03 & 318 \\
        218884  & 0.284 & 0.083 & 0.035 & 318 \\
        218889  & 0.284 & 0.083 & 0.035 & 140 \\
        219181  & 0.574 & 0.294 & 0.03 & 101 \\
        218897  & 1.38 & 0.082 & 0.03 & 140 \\
        219186  & 0.574 & 0.294 & 0.03 & 140 \\
        219197  & 0.574 & 0.294 & 0.03 & 49 \\
        218914  & 3.69 & 0.094 & 0.03 & 140 \\
        219207  & 0.574 & 0.294 & 0.03 & 140 \\
        218956  & 0.264 & 0.071 & 0.02 & 254 \\
        \hline
        \end{tabular}
    
    \label{tab:line_data}
    \end{table}
\end{center}

\subsubsection{Randomized scenario}

\begin{figure} [!ht] 
    \centering
    \includegraphics[height=8.5cm]{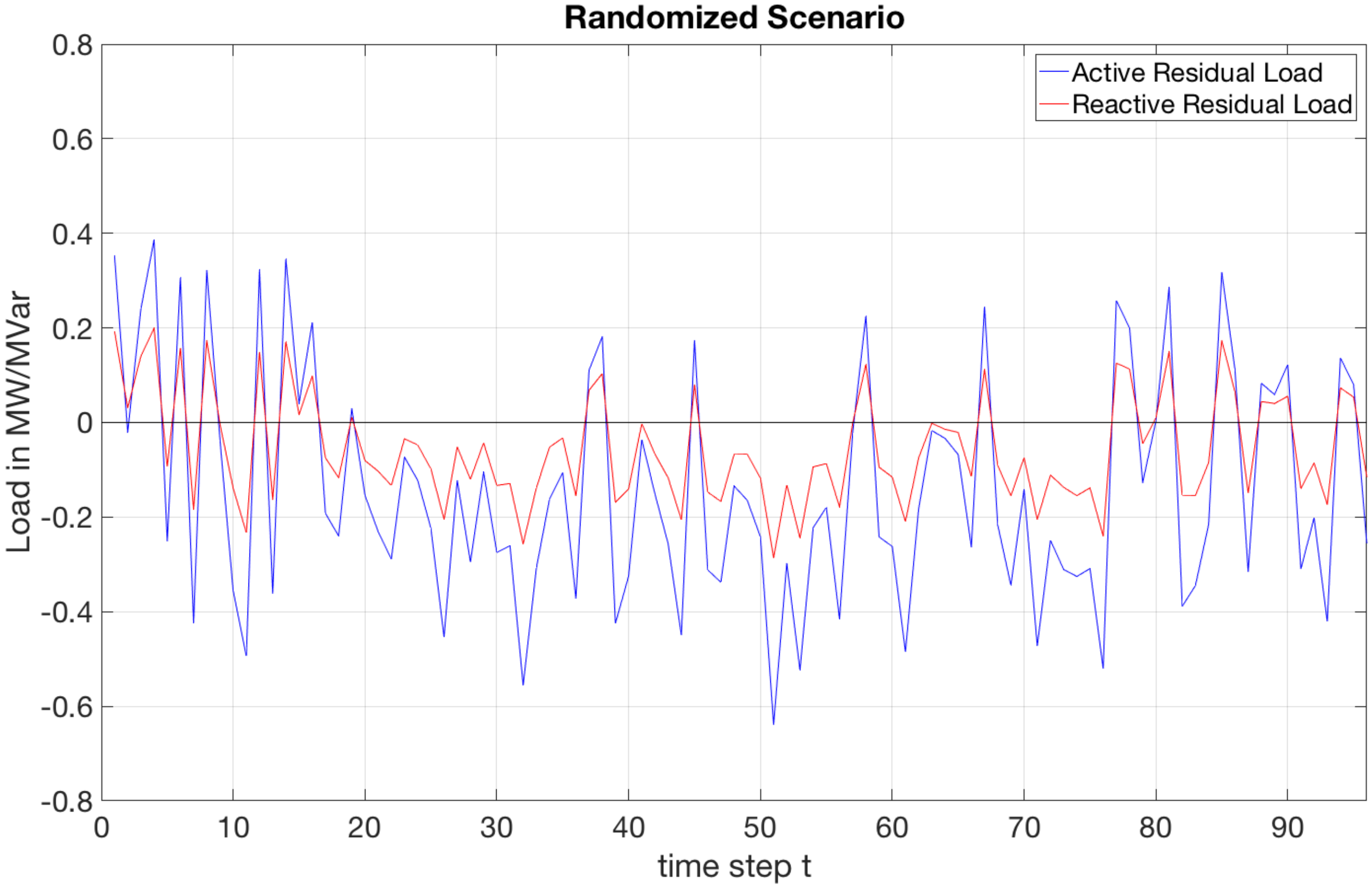}
    \caption{An exemplar residual load profile of our randomized simulation scenario for one day, that is, 96 time steps in 15 minute resolution.}
    \label{fig:randomized_scenario_residual}
\end{figure}

For the randomized scenarios, we assign fixed and randomized time series to wind generation and non-controllable consumption units. These time series are independently generated from a uniform distribution between 0 and 1, and are multiplied with a constant power rating. We then assign an individual and fixed time series to each consumption and generation unit that is sampled from an exponential distribution with the random time series values as expected value in each time step. The initial states of charge of each DR application is chosen from a uniform distribution between the upper and lower temperature bounds we introduce in our DR application models. Figure \ref{fig:randomized_scenario_residual} shows the active and reactive load profiles for one day, that is, 96 time steps with 15 minute resolution. 

\subsubsection{Regular scenario}

\begin{figure} [!ht] 
    \centering
    \includegraphics[height=8.5cm]{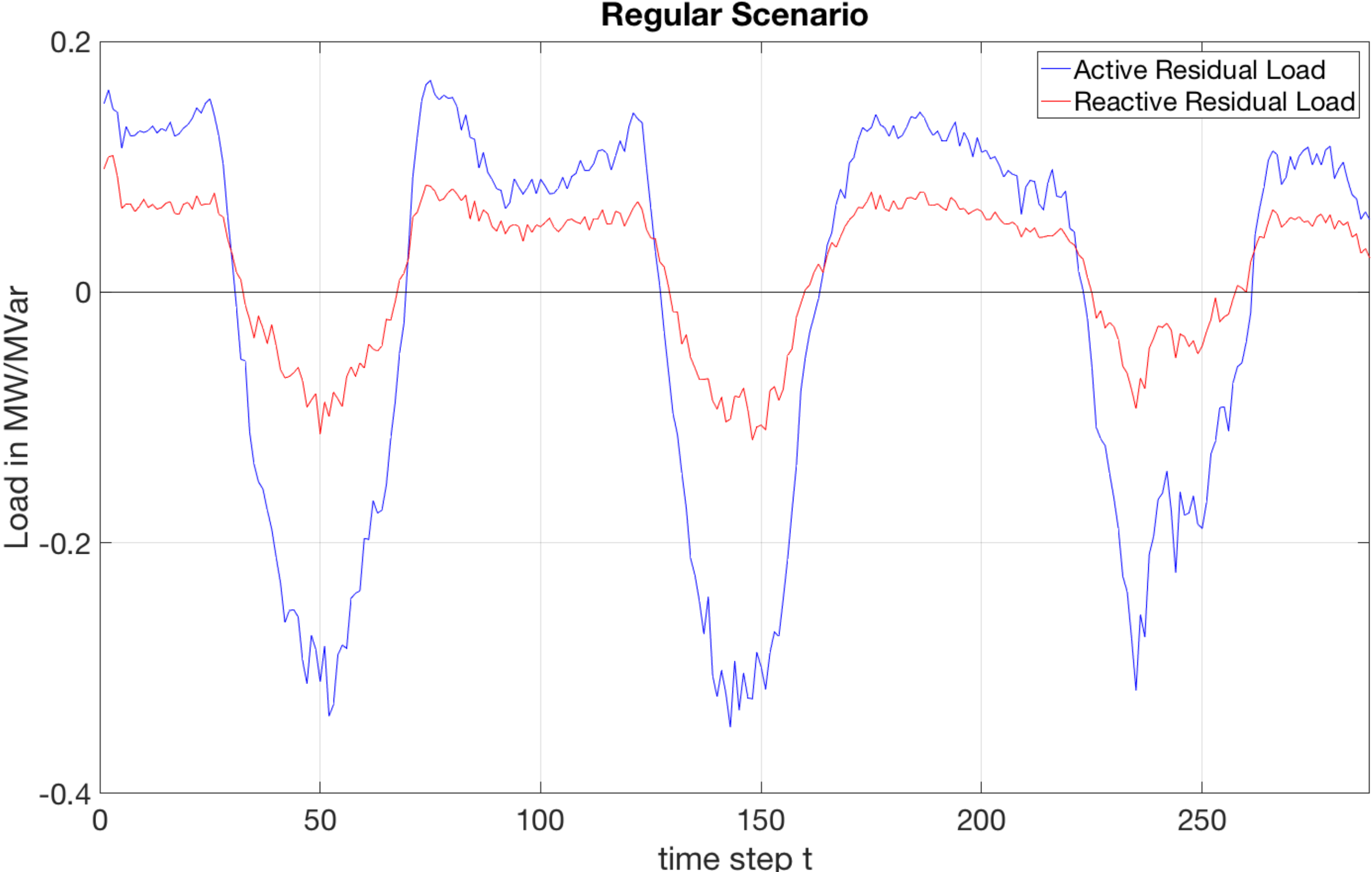}
    \caption{An exemplar residual load profile of our regular simulation scenario for three consecutive days, that is, 288 time steps in 15 minute resolution.}
    \label{fig:regular_scenario_residual}
\end{figure}

For the regular scenarios, we sample PV unit outputs from a Tennet 2010 generation profile, wind power unit outputs from a Tennet wind generation profile, and non-controllable loads from an ENTSO-E 2009 representative load profile, all from data that is made available in \emph{Adaptricity.sim} (\url{https://www.adaptricity.com/}). Figure \ref{fig:regular_scenario_residual} shows the resulting active and reactive residual load profiles for three consecutive days. 

\subsection{Simplified control objectives}

The net power injections over discrete time steps at each bus in the grid are represented by the residual load profiles at the bus, that is, the consumption of electricity minus the generation from renewable energy sources connected to this bus. If we assume that the residual load at every bus is zero, that means consumption and generation are perfectly matched at each bus, no power flows within the grid. As soon as the residual load somewhere in the grid deviates from zero, power demand flows into the bus and surplus generation flows out of it. The active and reactive flow of power $P_{km}/Q_{km}$ between two buses $k$ and $m$ is explained by the power flow equations as

\begin{align*}
    P_{km} & = U_k^2 g_{km} - U_k U_m g_{km} cos\theta_{km} - U_k U_m b_{km} sin\theta_{km} \\
    Q_{km} & = - U_k^2 (b_{km} + b_{km}^{sh}) + U_k U_m b_{km} cos\theta_{km} - U_k U_m g_{km} sin\theta_{km},
\end{align*}

where $U_{k}/U_{m}$ are voltage magnitudes at buses $k/m$, $\theta_{km}$ is the voltage angle difference between buses $k$ and $m$, $g_{km}$ is the real part of the admittance between $k$ and $m$, and $b_{km}$ is the imaginary part of the admittance between $k$ and $m$. We can read from these equations, that every flow of power between buses in the grid is associated with phase/power angle shifts, the relative timing between oscillating current and voltage, bringing the system \emph{out of step}. It is the reactive and capacitive properties of the lines and loads which make the current lead or lag the voltage in time. As the phase angle shifts, the magnitude of voltage also deviates from its nominal value \cite{Meier2006}. A dependence between active ($P_{km}^{line-loss}$) as well as reactive ($Q_{km}^{line-loss}$) power losses of a line and voltage drop ($|E_k - E_m|$) can be expressed as

\begin{align}
    P_{km}^{line-loss} & = P_{km} + P_{mk} =  g_{km} |E_k - E_m|^2 \\
    Q_{km}^{line-loss} & = Q_{km} + Q_{mk} = - b_{km}^{sh} (U_m^2 + U_k^2) - b_{km} |E_k - E_m|^2,
\end{align}

where $E_k/E_m$ are voltage phasors at buses $k/m$, and $b_{km}^{sh}$ is the imaginary part of the shunt element admittance between $k$ and $m$. In the Appendix, we provide a more comprehensive formulation of these relationships. Neglecting shunt elements and reformulating the voltage drop in dependence on power losses, we can deduce from equations (1) and (2) that:

\begin{align*}
    |E_k - E_m| & \sim \sqrt{P_{km}^{line-loss}/g_{km}}  \\
    |E_k - E_m| & \sim \sqrt{-Q_{km}^{line-loss}/b_{km}}
\end{align*}

We can observe that the voltage deviation between two buses increases with the losses associated with power flowing along their connecting line, as line admittances are constant. We further know that controlling the residual load at buses, that is, the net power injections over time, allows us to coordinate bi-directional power flows. Given the relationships derived above, we can yield that:

\begin{enumerate}
    \item Reducing the path along which power flows in a grid reduces overall voltage deviations.
    \item Reducing peak power flows along a line reduces peak voltage deviations.
\end{enumerate}

This motivates us to examine two simplified control objectives for coordinating bi-directional power flows and regulating voltage deviations with DR:

\begin{enumerate}
    \item Reducing the total power exchange of a grid with the remaining system.
    \item Reducing the peak power flows along all lines in a grid.
\end{enumerate}

Compared to traditional methods in power systems, the first objective alleviates the need for any information about the topology of a grid and its state at each bus. In contrast, the second objective attempts to leverage this data in order to create better-informed control decisions.

\subsection{Mathematical optimization}
Here, we formulate the mathematical optimization problems that must be solved for achieving the simplified objectives we formulate above. We introduce two problem formulations which we incorporate into our DR control algorithms later on. The first problem is generally applicable and handles all occurring situations with a quadratic formulation. The second problem formulation is a linear approximation that leads to faster computational time but is only applicable to special cases. 

\subsubsection{General quadratic objective}
Based on a prediction of the non-controllable residual load $r_t$ over time steps $t = 1 \dots T$, we want to schedule flexible loads such that positive and negative peaks of the resulting total residual load are reduced. Binary decisions on the external device switches $u_t$ determine when devices have to be turned on and off. A quadratic formulation of this objective is given as

\begin{equation*}
    \begin{aligned}
        & \underset{u}{\text{min}}
        & & \sum\nolimits_{t=1}^T (r_t + P   u_t)^2\\
    \end{aligned}
\end{equation*}

\begin{figure}[ht]
    \centering
    \includegraphics[height = 9cm]{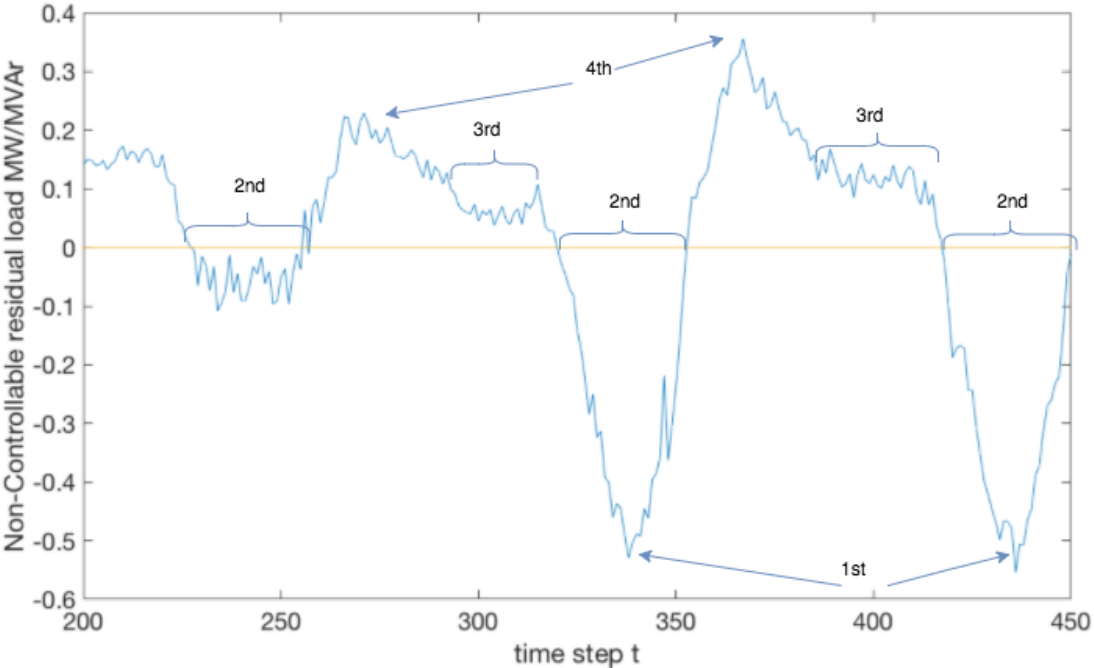} 
    \caption{A visualization of the scheduling strategy when using a general quadratic objective function for positive and negative peak shaving of residual load. The blue line represents the non-controllable residual load. The numbers describe the order in which the optimizer will activate flexible loads for peak shaving.}
    \label{fig:optimizer_1}
\end{figure}

This objective realizes positive and negative peak shaving. Figure \ref{fig:optimizer_1} visualizes how flexible loads are scheduled, consisting of the following order of priority:

\begin{enumerate}
    \item negative peaks of r\textsubscript{t}
    \item negative r\textsubscript{t}
    \item positive r\textsubscript{t}
    \item positive peaks of r\textsubscript{t} 
\end{enumerate}

\subsubsection{Approximate linear objective}

We can improve the computational time for solving the above objective using an approximate linear formulation of the objective. This can be formulated as

\begin{equation*}
    \begin{aligned}
        & \underset{u}{\text{min}}
        & & \sum\nolimits_{t=1}^T |r_t - r_{t-1} + P   u_t|\\
    \end{aligned}
\end{equation*}

\begin{figure}[ht]
    \centering
    \includegraphics[height = 9cm]{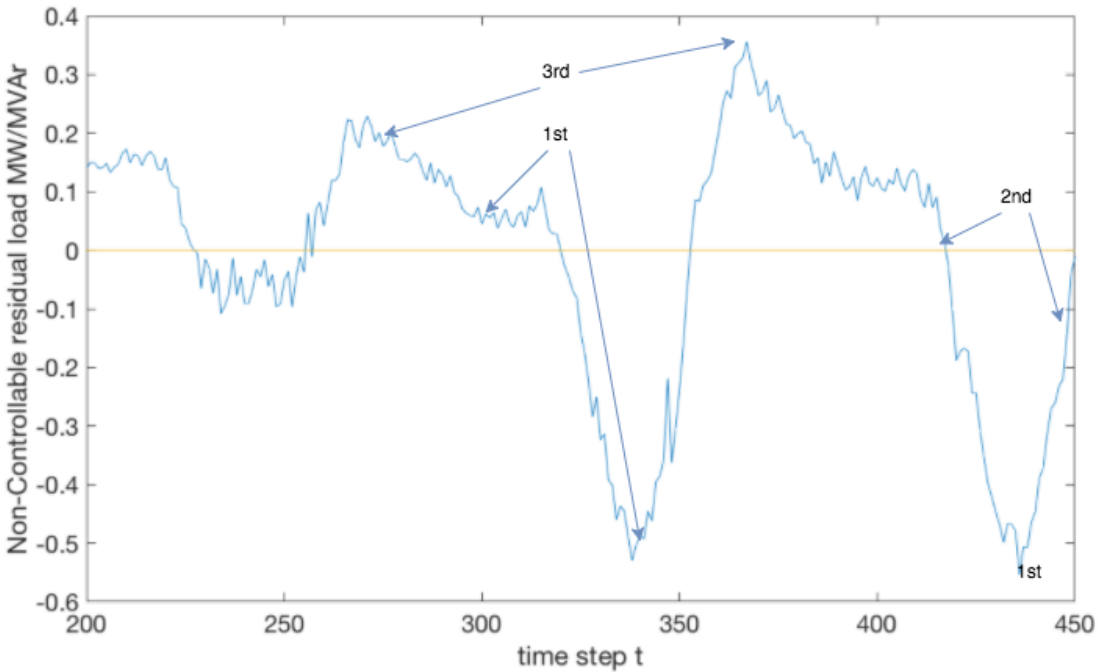} 
    \caption{A visualization of the scheduling strategy when using an approximate linear objective function for allocating DR resources. The blue line represents non-controllable residual load. The numbers indicate the order in which the optimizer prefers to activate flexible loads.}
    \label{fig:optimizer_2}
\end{figure}

This reduces the number of different scheduling situations from the initial four to three and makes it easier for the optimizer to find a solution. Figure \ref{fig:optimizer_2} visualizes how flexible loads are scheduled. The approximate formulation reduces the sum over the magnitude of gradients of the total residual load by activating flexible loads. DR applications with a relatively small shifting capacity, for example fridges, are strongly restricted in the time range over which they can be scheduled without causing consumer comfort loss. For scenarios with long periods between negative and positive non-controllable residual load $r_t$, and for flexible loads with small time constants, this leads to a similar scheduling as with the quadratic optimizer. The priority order in which the optimizer schedules flexible loads is as follows:

\begin{enumerate}
    \item valleys of positive or negative $r_t$ 
    \item slope of positive or negative $r_t$
    \item peak of positive or negative $r_t$
\end{enumerate}

\subsubsection{Constraints}
We can formulate constraints in the same way for both formulations of the objective function. We can mainly write these as

\begin{equation*}
    \begin{aligned}
    &x_{t+1}  && =  x_t - x_{t}^{out} - x^{iso}_t + x^{inp}_t   &&& \forall t = 0 \dots T-1  \\
    &T_{low}  && \leq   x_t  \leq \ T_{up}   &&& \forall t = 0 \dots T-1 \\
    &(T_{low}+T_{up})/2  && \leq  x_N  \leq   T_{up}    &&& \forall t = T \\
    &u_t \in \{0,1\} && &&& \forall t = 0 \dots T-1 \\
    \end{aligned}
\end{equation*}

We are considering the iterative change of internal storage states in the equality constraint. The internal temperatures are kept within upper (T\textsubscript{up}) and lower (T\textsubscript{low}) bounds in order to maintain consumer comfort. In the last time step N, the internal storage state $x_N$ is requested to be at least half full so as to avoid that the optimizer delivers a completely empty storage. 

Subsequently, we reformulate the constraints of our problem in pure dependence on the decision variables by substituting the internal state temperature. This allows us to better understand the mathematical properties of the problem we are trying to solve. Using mathematical induction, we find that a generally applicable expression is

\begin{align*}
    x_t & = f(x_0, u_1, \dots , u_t) \\
        & = x_0 (1- c_{los})^{t+1} + \sum\nolimits_{n=0}^t (c_{los} x_n + x_n^{inp} - x_n^{iso} - x_n^{out} )   (1 - c_{los})^{t-n} 
\end{align*}

We see that the internal temperature in each time step is dependent on decisions made in all earlier time steps. Note that the decision variable u\textsubscript{t} is contained in $x_t^{inp}$. For each of the DR applications we model, we can hence deduce the following equations.

\textbf{Storage water boiler}

\begin{equation*}
    \begin{split}
        t=0: \\
        x_1 & = x_0 - c_{use} s_{3,0} - c_{los} (x_0 - T_{ss}) +c_{inp} P u_0 \\
            & = x_0 (1 - c_{los}) - c_{use} s_{3,0} + c_{los} T_{ss} +c_{inp} P u_0\\
        t=1: \\
        x_2 & = x_1 (1 - c_{los}) - c_{use} s_{3,1} + c_{los} T_{ss} + c_{inp} P u_1 \\
            & = x_0 (1 - c_{los})^2 +(-c_{use} s_{3,0} + c_{los} T_{ss} + c_{inp} P u_0) (1 - c_{los}) - c_{use} s_{3,1} + c_{los} T_{ss} + c_{inp} P u_1 \\
        \forall t \geq 0: \\
        x_t & = x_0 (1- c_{los})^{t+1} + \sum\nolimits_{n=0}^t (c_{los} T_{ss} + c_{inp} P u_n - c_{use} s_{3,n}) (1 - c_{los})^{t-n} 
    \end{split}
\end{equation*}

\textbf{Night storage heater}

\begin{equation*}
    \begin{split}
        t=0: \\
        x_1 & = x_0 - max(0, c_{use} (T_{ss} - s_{1,0}) - c_{sol} s_{2,0}) - c_{los} (x_0 - T_{ss}) + c_{inp} P u_0 \\
            & = x_0 (1 - c_{los}) - max(0, c_{use} (T_{ss} - s_{1,0}) - c_{sol} s_{2, 0}) + c_{los} T_{ss} + c_{inp} P u_0\\
        t=1: \\
        x_2 & = x_1 (1 - c_{los}) - max(0, c_{use} (T_{ss} - s_{1,1}) - c_{sol} s_{2, 1}) + c_{los} T_{ss} + c_{inp} P u_1\\
            & = x_0 (1 - c_{sol})^2 + (1 - c_{los}) (-max(0, c_{use} (T_{ss} - s_{1, 0}) - c_{sol} s_{2, 0}) + c_{los} T_{ss} + c_{inp} P u_0) \\
            & - max(0, c_{use} (T_{ss} - s_{1, 1}) - c_{sol} s_{2, 1}) + c_{los} T_{ss} + c_{inp} P u_1  \\
        \forall t \geq 0: \\
        x_t & = x_0 (1 - c_{los})^{t+1} + \sum\nolimits_{n=0}^t (c_{los} T_{ss} + c_{inp} P u_n - max(0, c_{use} (T_{ss} - s_{1,n}) -  c_{sol} s_{2,n})) (1-c_{los})^{t-n} 
    \end{split}
\end{equation*}

\textbf{Heat pump}

\begin{equation*}
    \begin{split}
        t=0: \\
        x_1 & = x_0 - max(0, c_{11} (T_{ss} - s_{1,0}) - c_{sol} s_{2,0}) - c_{12} s_{3,0} - c_{los} (x_0 - T_{ss}) \\
            & + c_{inp} P (1- \frac{max(0,min(s_{1,0} - T_{LOL}, T_{HOL} - T_{LOL}))} {T_{HOL} - T_{LOL}} )u_0\\
            & = x_0 (1 - c_{los}) - max(0, c_{11} (T_{ss} - s_{1,0}) - c_{sol} s_{2,0}) - c_{12} s_{3,0} + c_{los} T_{ss}  \\
            & + c_{inp} P (1- \frac{max(0,min(s_{1,0} - T_{LOL}, T_{HOL} - T_{LOL}))} {T_{HOL} - T_{LOL}} )u_0\\
        t=1: \\
        x_2 & = x_1 (1 - c_{los}) - max(0, c_{11} (T_{ss} - s_{1,1}) - c_{sol} s_{2,1}) - c_{12} s_{3,1}     + c_{los} T_{ss} \\
            & + c_{inp} P (1- \frac{max(0,min(s_{1,1} - T_{LOL}, T_{HOL} - T_{LOL}))} {T_{HOL} - T_{LOL}}) u_1\\
        \forall t \geq 0: \\
        x_t & = x_0 (1 - c_{los})^{t+1} +  (1 - c_{los})^{t - n}   \sum\nolimits_{n=0}^t (c_{los} T_{ss} - c_{12} s_{3,n} + c_{inp} P \\
        & \cdot (1- \frac{max(0,min(s_{1,n} - T_{LOL}, T_{HOL} - T_{LOL}))} {T_{HOL} - T_{LOL}} ) u_n - max(0, c_{11} (T_{ss} - s_{1,n}) - c_{sol} s{2,n})) 
    \end{split}
\end{equation*}

\textbf{Fridge and Freezer}

\begin{equation*}
    \begin{split}
        t=0: \\
        x_1 & = x_0 + c_{use} s_{4,0} + c_{los} (T_{ss} - x_0) - c_{inp} P u_0\\
            & = x_0 (1 - c_{los}) + c_{use} s_{4,0} + c_{los} T_{ss} - c_{inp} P u_0 \\
        t=1: \\
        x_2 & = x_1 (1 - c_{los}) + c_{use} s_{4,1} + c_{los} T_{ss} - c_{inp} P u_0\\
            & = x_0 (1 - c_{los})^2 + (c_{use} s_{4,0} + c_{los} T_{ss} - c_{inp} P u_0) (1 - c_{los}) + c_{use} s_{4,1} + c_{los} T_{ss} + c_{inp} P u_1 \\
        \forall t \geq 0: \\
        x_t & =  x_0 (1 - c_{los})^{t+1} + \sum\nolimits_{n=0}^t (c_{use} s_{4,n} + c_{los} T_{ss} - c_{inp} P u_n) (1-c_{los})^{t- n} 
    \end{split}
\end{equation*}

\textbf{Air conditioner}

\begin{equation*}
    \begin{split}
        t=0: \\
        x_1 & = x_0 + c_{use} s_{4,0} + c_{los} (s_{1,0} - x_0) + c_{sol} s_{2,0} - c_{inp} P u_0\\
            & = x_0 (1 - c_{los}) + c_{use} s_{4,0} + c_{los} s_{1,0} + c_{sol} s_{2,0} - c_{inp} P u_0 \\
        t=1: \\
        x_2 & = x_1 (1 - c_{los}) + c_{use} s_{4,1} + c_{los} s_{1,1} + c_{sol} s_{2,1} - c_{inp} P u_1 \\
            & = x_0 (1 - c_{los})^2 + (c_{use} s_{4,0} + c_{los} s_{1,0} + c_{sol} s_{2,0} - c_{inp} P u_0) (1 - c_{los}) \\ 
            & + c_{use} s_{4,1} + c_{los} s_{1,1} + c_{sol} s_{2,1} - c_{inp} P u_1 \\
        \forall t \geq 0: \\
        x_t & =  x_0 (1 - c_{los})^{t+1} + \sum\nolimits_{n=0}^t (c_{use} s_{4,n} + c_{los} s_{1,n} + c_{sol} s_{2,n} - c_{inp} P u_n) (1-c_{los})^{t- n} 
    \end{split}
\end{equation*}

\subsubsection{Problem properties}

We can relax the constraints of our problem with a factor $a_t$ while ensuring that a solution is found. In the following, we describe six categories of choice for the problem that we want to solve. Several heuristics and approximations can lead to faster computational time.

\textbf{Objective function}. Different formulations of the objective function can be derived from the general p-Norm:

\begin{equation*}
    || x ||_p =  (\sum\nolimits_{t=1}^T |x_t|^p)^{\frac{1}{p}}
\end{equation*}

The formulation of the objective function is important to the results, depending on which solver one uses. Some solvers can use an Euclidean formulation to transform the quadratic problem into a second-order-cone problem and apply alternative algorithms that can lead to a faster solution of the problem. For the approximate linear problem only the formulation we derived above applies.

\begin{equation*}
    \begin{split}
        \text{Sum-Norm \hspace{4mm}} & \underset{u,a}{\text{min}} \quad \sum\nolimits_{t=1}^T (r_t + P   u_t)^2 +a_t \\
        \text{Euclidean-Norm \hspace{4mm}} & \underset{u,a}{\text{min}} \quad || r + P   u ||_2 + \sum\nolimits_{t=1}^T a_t \\
        \text{Maximum-Norm \hspace{4mm}} & \underset{u,a}{\text{min}} \quad || r + P   u ||_\infty + \sum\nolimits_{t=1}^T a_t \\
    \end{split}
\end{equation*}

\textbf{Power optimization}. Each DR application is assigned with a load factor smaller than one, such that we also have control over reactive power flows. We can therefore choose between pure active, pure reactive or both active and reactive power flow optimization. This applies to both the general and the approximate formulation.

\begin{equation*}
    \begin{split}
        \text{active \hspace{4mm}} &  \underset{u,a}{\text{min}} \quad \sum\nolimits_{t=1}^T (r_t^{act} + P   u_t)^2 +a_t \\
        \text{reactive \hspace{4mm}} &  \underset{u,a}{\text{min}} \quad \sum\nolimits_{t=1}^T (r_t^{react} + Q   u_t)^2 +a_t \\
        \text{active \& reactive \hspace{4mm}} & \underset{u,a}{\text{min}} \quad \sum\nolimits_{t=1}^T (r_t^{act} + P   u_t)^2 + (r_t^{react} + Q   u_t)^2 +a_t \\
    \end{split}
\end{equation*}

\textbf{Internal controller}. The internal controller plays an important role for solving our control problem in the real world. We can choose whether full control via external signals over the DR application is assumed or whether an internal controller is considered. In both cases, we introduce a weighting factor $w_1$ which is used to prioritize the importance of consumer comfort. If the internal controller is considered, we can additionally tune a weighting factor $w_2$ which prioritizes to what extent the optimizer should stay away from certain margins, such as the dead-band temperatures of the internal storage of a thermal DR application. The internal controller for a heating device can be formulated as 

\begin{equation*}
    \begin{split} 
        \text{full control \hspace{4mm}} & T_{low} - \frac{1}{w_1} a_t \leq x_t \leq T_{up}  + \frac{1}{w_1} a_t \\
        \text{internal controller \hspace{4mm}} & T_{low}  - \frac{1}{w_1} a_t \leq x_t \leq T_{up} - T_{DB} + \frac{1}{w_2} a_t \\
    \end{split}
\end{equation*}

\textbf{Decision variables}. Originally, we pursue binary signals (on = 1, off = 0) for the optimization. Depending on the solver, the problem can be solved much faster if we formulate the problem with non-integer variables between 0 and 1 and round the solution to binary afterwards. Another approach is to simulate in time steps that are rougher than the signals that are actually applied such that solutions between 0 and 1 can be rounded into binary signals for each time step of application. The average consumption over the calculated time step will hence stay the same as we can distribute binary signals over this rougher time step. This further has the advantage of faster computation at the cost of less accurate control signal actions.

\begin{equation*}
    \begin{split}
        \text{mixed-integer \hspace{4mm}} &  a_t \in \mathbb{R}_{\geq 0},  u_t \in \{0,1\}   \\
        \text{integer \hspace{4mm}}  &  a_t \in \mathbb{N}_{0},  u_t \in \{0,1\}   \\
        \text{non-integer \hspace{4mm}}  &  a_t \in \mathbb{R}_{\geq 0}, u_t \in [0,1]   \\
    \end{split}
\end{equation*}

\textbf{Sequential solutions}. The optimization can be executed for one DR application at the time, or, for a batch of applications simultaneously. By splitting one large problem for all DR applications in a grid into a sequence of sub-problems for each DR application in a grid, the complexity of our optimization can be significantly reduced. A trade-off is likely paid in the accuracy of the results.

\subsection{Demand Response control algorithms}

We develop DR control algorithms that incorporate the optimization of the above tasks. This allows us to achieve our simplified objectives. For each objective, we introduce two algorithms. We call these the \emph{optimal/heuristic-grid-algorithm} and the \emph{optimal/heuristic-line-algorithm}. The optimal-grid/line-algorithms dispatch all controllable devices in the grid in one large optimization problem with high complexity. The heuristic-grid/line-algorithms instead divide the problem into significantly smaller sub-problems by scheduling only one device at a time. By doing this, the complexity of our optimization is reduced at the price of less accurate solutions. Therefore, we name the latter two algorithms heuristic. A fifth algorithm, which we call the \emph{critical-line-algorithm}, further combines the concepts of the first two approaches and is designed to treat power flows over specified critical lines with higher priority. In the following, we provide an exemplar problem formulation for each algorithm by making explicit choices about the properties of our problem.

\subsubsection{Optimal-grid-algorithm}

\begin{figure}[ht]
    \centering
    \includegraphics[height = 7.5cm]{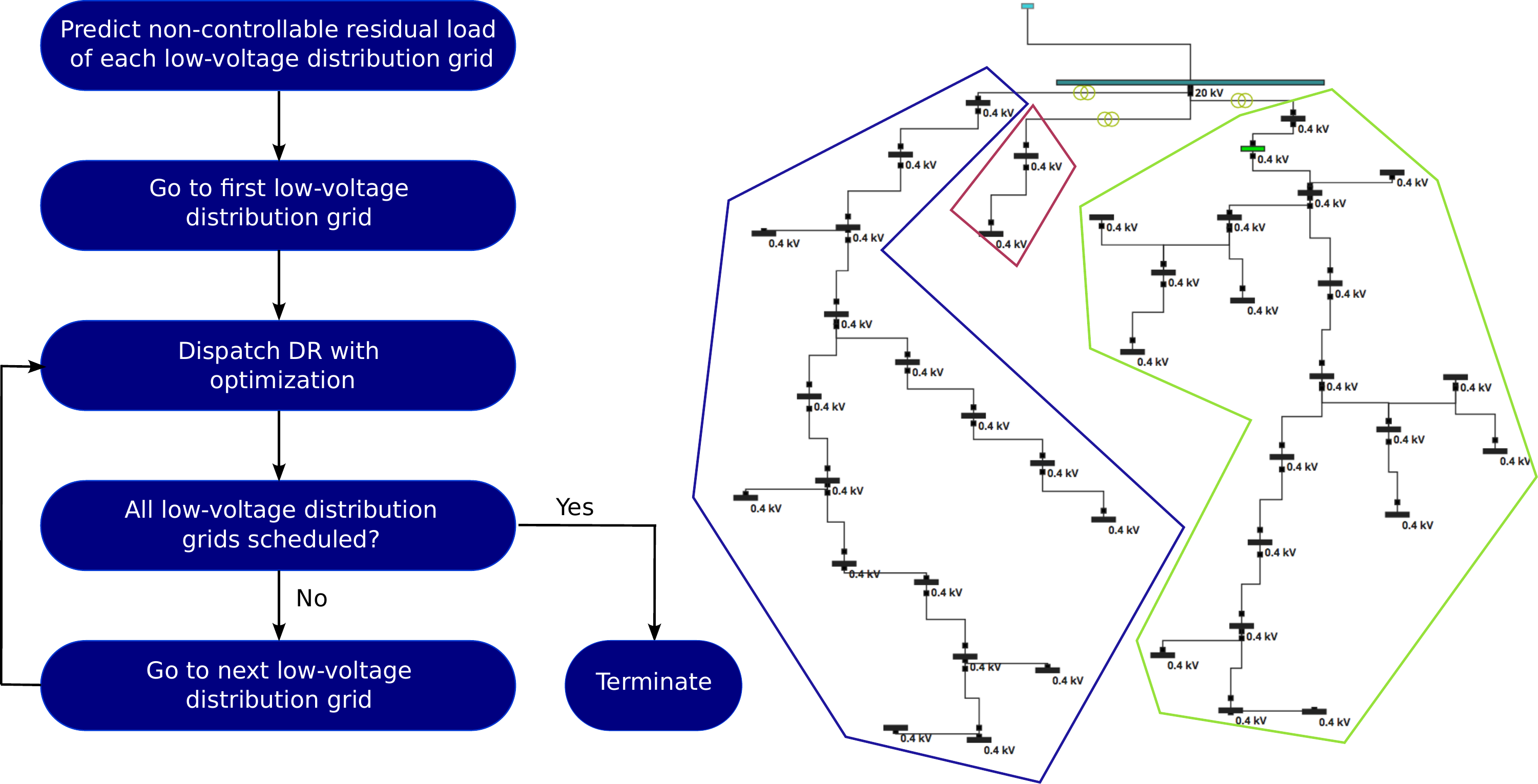} 
    \caption{Flow chart for optimal-grid-algorithm.}
    \label{fig:optimal_grid}
\end{figure}

The optimal-grid-algorithm minimizes the peak power exchange of each low-voltage grid with the medium voltage grid separately. It minimizes the peak load at each transformer sub-station, and therefore pursues our first simplified control objective. First, the non-controllable residual load in an entire grid is assumed to be predicted. Then, one optimization problem dispatches all devices of the same grid at the same time. Figure \ref{fig:optimal_grid} provides a flow chart of this algorithm. A quadratic formulation with Euclidean objective, active and reactive power optimization, relaxed constraints, full control and integer decision variables can be written as:

\begin{align*}
    & \underset{u_t, a_t}{\text{min}} || r^{act} + \sum\nolimits_{i=1}^I P_i   u_{i} ||_2 + || r^{react} + \sum\nolimits_{i=1}^I Q_i   u_{i} ||_2 + \sum\nolimits_{t=1}^T \sum\nolimits_{i=1}^I a_{i,t}    \\
    & \text{s.t.}   \\
    & T_{low,i} - \frac{a_{i,1}}{w_1} \leq x_{i,0}   (1 - c_{i,3})^{1+1} + \sum\nolimits_{n=0}^1 (c_{i,3} x_{i,n} + x_{i,n}^{inp} - x_{i,n}^{ins} - x_{i,n}^{out})    \\ & \cdot (1 - c_{i,3})^{1-n} \leq T_{up,i} + \frac{a_{i,1}}{w_1} \qquad \forall i = 1 \dots I   \\
    & T_{low,i} - \frac{a_{i,2}}{w_1} \leq x_{i,0}   (1 - c_{i,3})^{2+1} + \sum\nolimits_{n=0}^2 (c_{i,3} x_{i,n} + x_{i,n}^{inp} - x_{i,n}^{ins} - x_{i,n}^{out})   \\ & \cdot (1 - c_{i,3})^{2 - n} \leq T_{up,i} + \frac{a_{i,2}}{w_1} \qquad \forall i = 1 \dots I     \\
    & \qquad \qquad \qquad \qquad \qquad \qquad \qquad \vdots  \\
    & \frac{T_{low,i} + T_{up,i}}{2} - \frac{a_{i,T}}{w_1} \leq x_{i,0}   ( 1-c_{i,3})^{T+1} + \sum\nolimits_{n=0}^T (c_{i,3} x_{i,n} + x_{i,n}^{inp} - x_{i,n}^{ins} - x_{i,n}^{out})   \\ & \cdot (1 - c_{i,3})^{T -n} \leq T_{up,i} + \frac{a_{i,T}}{w_1} \qquad \forall i = 1\dots I    \\
    & u_{i,t} \in \{ 0,1 \} \quad, \quad a_{i,t} \in \mathbb{N}_0 \qquad \forall i = 1 \dots I, \forall t = 1 \dots T
\end{align*}

\subsubsection{Heuristic-grid-algorithm}

\begin{figure}[ht]
    \centering
    \includegraphics[height = 8cm]{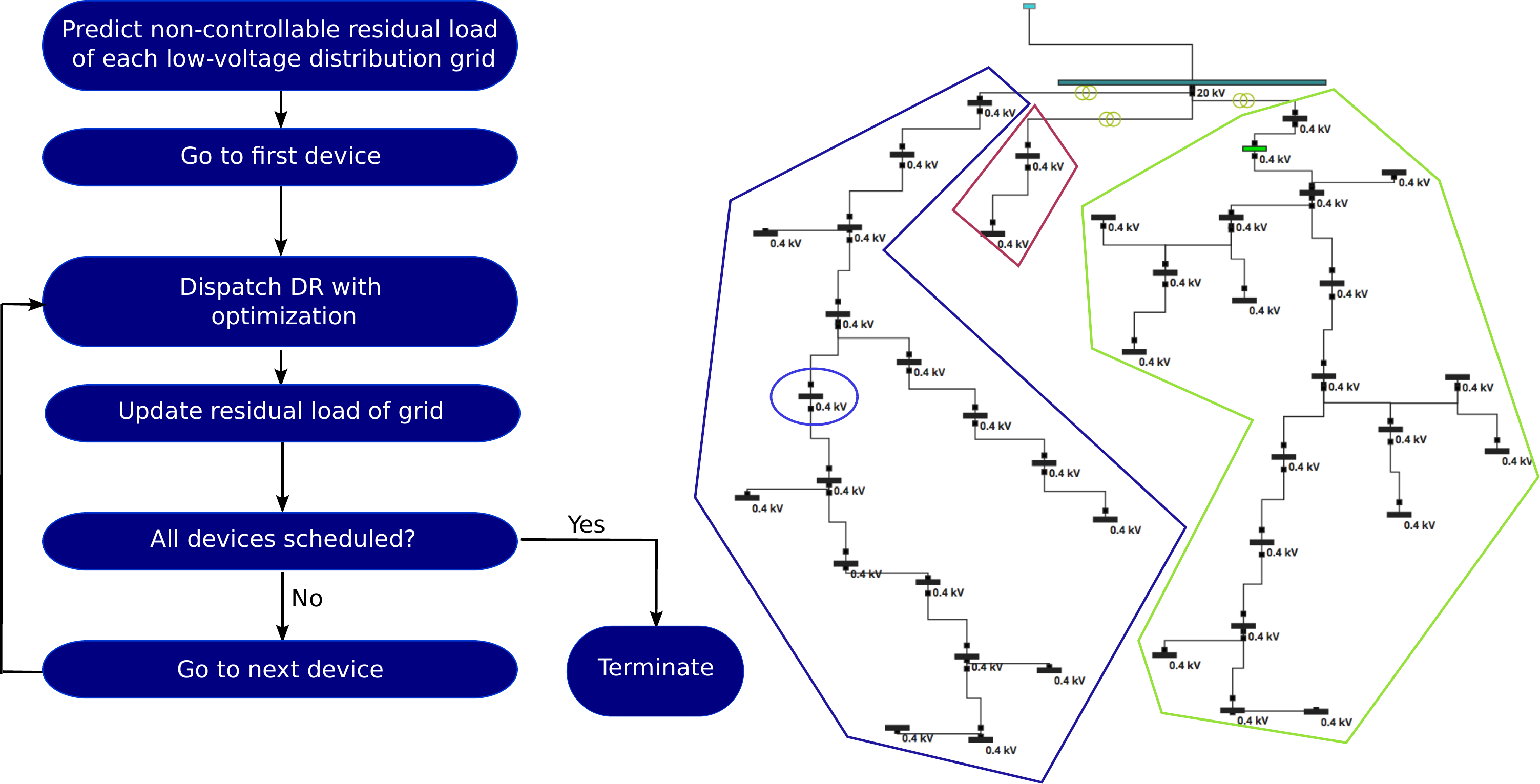} 
    \caption{Flow chart for heuristic-grid-algorithm.}
    \label{fig:heuristic_grid}
\end{figure}

The heuristic-grid-algorithm reduces the peak power exchange of each low-voltage grid with the medium voltage grid. It reduces the peak load at each transformer sub-station, and hence also pursues our first simplified control objective. The main difference to the optimal-grid-algorithm is that peak residual loads are not reduced optimally anymore as devices are dispatched one after another in separate optimization problems. Again, the non-controllable residual load of an entire grid is assumed to be predicted first. Then, each device is dispatched by solving an own optimization problem for its associated grid. In-between, the residual load of the associated grid $r_t$ is updated by adding the dispatched load profiles to the non-controllable residual load of the grid. Figure \ref{fig:heuristic_grid} provides a flow chart of the algorithm. An exemplary formulation with a Sum-Norm quadratic objective, reactive power optimization, relaxed constraints, full control and non-integer decision variables can be written as:

\begin{align*}
    & \underset{u_t, a_t}{\text{min}} \sum\nolimits_{t=1}^T (r_t^{react} + Q   u_t)^2 + a_t    \\
    & \text{s.t.}    \\
    & T_{low} - \frac{a_1}{w_1} \leq x_0   (1 - c_{los})^{1+1} + \sum\nolimits_{n=0}^1 (c_{los} x_n + x_n^{inp} - x_n^{ins} - x_n^{out})   \\ & \cdot (1 - c_{los})^{1-n} \leq T_{up} + \frac{a_1}{w_1}  \\
    & T_{low} - \frac{a_2}{w_1} \leq x_0   (1 - c_{los})^{2+1} + \sum\nolimits_{n=0}^2 (c_{los} x_n + x_n^{inp} - x_n^{ins} - x_n^{out})   \\ & \cdot (1 - c_{los})^{2 - n} \leq T_{up} + \frac{a_2}{w_1}    \\
    & \qquad \qquad \qquad \qquad \qquad \qquad \qquad \vdots  \\
    & \frac{T_{low} + T_{up}}{2} - \frac{a_T}{w_1} \leq x_0   ( 1-c_{los})^{T+1} + \sum\nolimits_{n=0}^T (c_{los} x_n + x_n^{inp} - x_n^{ins} - x_n^{out})   \\ & \cdot (1 - c_{los})^{T -n} \leq T_{up} + \frac{a_T}{w_1}   \\
    & u_t \in [0,1] \quad, \quad a_t \in \Re_{\geq 0} \qquad \forall t = 1 \dots T  
\end{align*}

\subsubsection{Optimal-line-algorithm}

\begin{figure}[ht]
    \centering
    \includegraphics[height = 8cm]{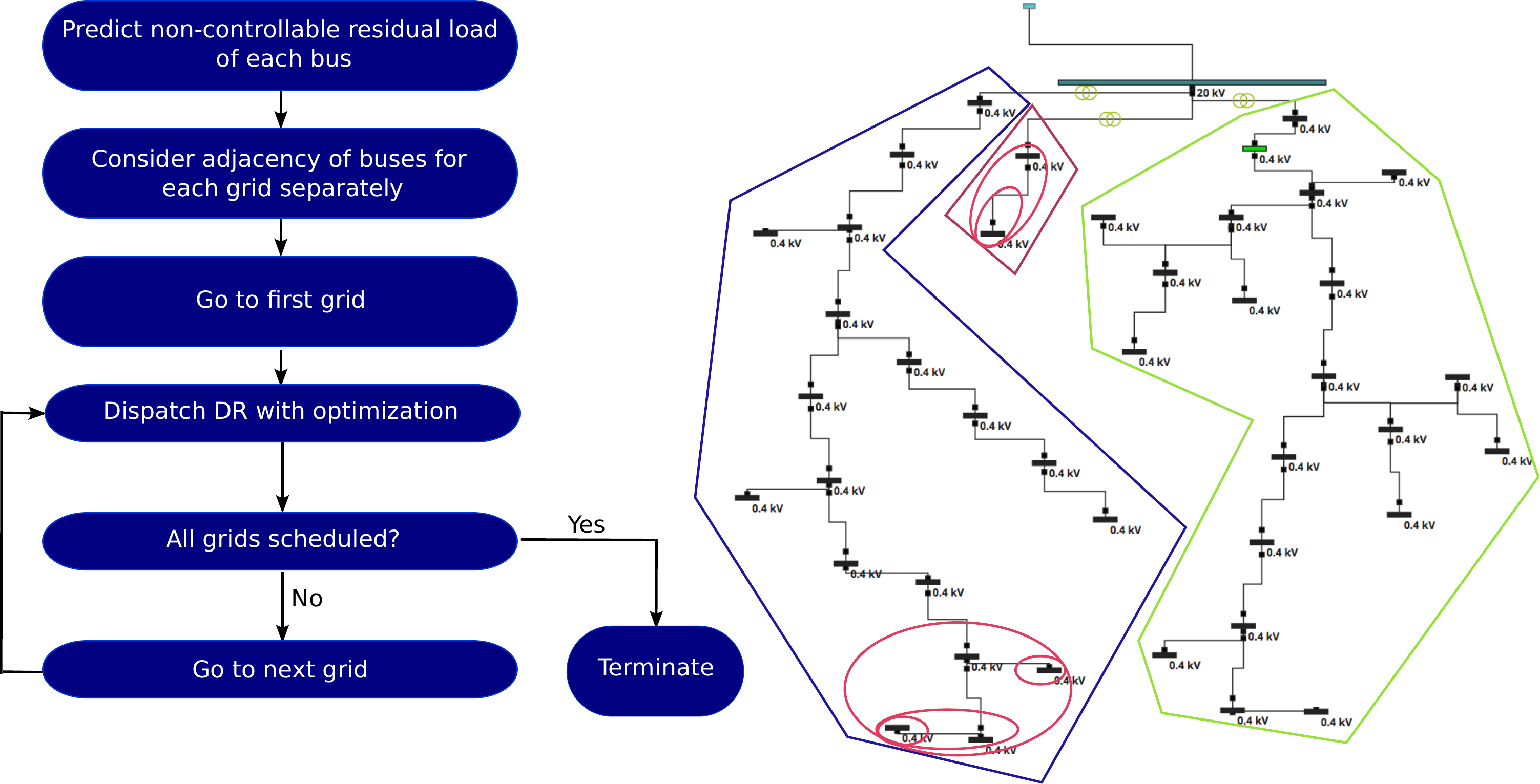} 
    \caption{Flow chart for optimal-line-algorithm.}
    \label{fig:optimal_line}
\end{figure}

The optimal-line-algorithm minimizes the sum of peak power flow along each line in the grid. For this, it must consider the topology of the grid. We start with predicting the non-controllable residual load $r_{i}$ at each bus $i$, where $i = 1 \dots I$. Now, we formulate the adjacency and power flows along each line in equality constraints: a first set of equality constraints sets the total residual load $P_i$ at bus $i$ equal to the non-controllable residual load $r_i$ at that bus plus all $K$ flexible loads $P_{i, k}$ connected to that bus; a second set of equality constraints formulates the power flow through line $i$ as the sum of the residual loads of all $J$ buses connected between line $i$ and the end of its branch, that is, all buses towards end of laterals, away from the feeding transformer. Figure \ref{fig:optimal_line} provides a flow chart of the algorithm. An exemplar formulation with quadratic objective, active power optimization, relaxed constraints, full control and mixed-integer decision variables can be written as:

\begin{align*}
    & \underset{u_t, a_t}{\text{min}} \sum\nolimits_{t=1}^T \sum\nolimits_{i=1}^I (P_{i,t}^{flow})^2  + a_t    \\
    & \text{s.t.}    \\
    & P_i = r_{i}^{act} + \sum\nolimits_{k=1}^K P_{i,k}   u_{k}    \qquad  \forall i=1 \dots  I     \\
    & P_i^{flow} =  \sum\nolimits_{j=1}^J P_{j}    \qquad  \qquad \qquad \forall i=1 \dots  I     \\
    & T_{low,k} - \frac{a_{k,1}}{w_1} \leq x_{k,0}   (1 - c_{k,3})^{1+1} + \sum\nolimits_{n=0}^1 (c_{k,3} x_{k,n} + x_{k,n}^{inp} - x_{k,n}^{ins} - x_{k,n}^{out})    \\ & \cdot (1 - c_{k,3})^{1-n}  \leq T_{up,k} + \frac{a_{k,1}}{w_1} \qquad \forall k = 1\dots K   \\
    & T_{low,k} - \frac{a_{k,2}}{w_1} \leq x_{k,0}   (1 - c_{k,3})^{2+1} + \sum\nolimits_{n=0}^2 (c_{k,3} x_{k,n} + x_{k,n}^{inp} - x_{k,n}^{ins} - x_{k,n}^{out})    \\ &  \cdot (1 - c_{k,3})^{2 - n}  \leq T_{up,k} + \frac{a_{k,2}}{w_1} \qquad \forall k = 1\dots K    \\
    & \qquad \qquad \qquad \qquad \qquad \qquad \qquad \vdots  \\
    & \frac{T_{low,k} + T_{up,k}}{2} - \frac{a_{k,T}}{w_1} \leq x_{k,0}   ( 1-c_{k,3})^{T+1} + \sum\nolimits_{n=0}^T (c_{k,3} x_{k,n} + x_{k,n}^{inp} - x_{k,n}^{ins} - x_{k,n}^{out})    \\ & \cdot (1 - c_{k,3})^{T -n} \leq T_{up,k} + \frac{a_{k,T}}{w_1} \qquad \forall k = 1\dots K    \\
    & u_{k,t} \in [0,1], \quad a_{k,t} \in N_0 \qquad \forall k = 1\dots K \qquad \forall t = 1\dots T  
\end{align*}

\subsubsection{Heuristic-line-algorithm}

\begin{figure}[ht]
    \centering
    \includegraphics[height = 7cm]{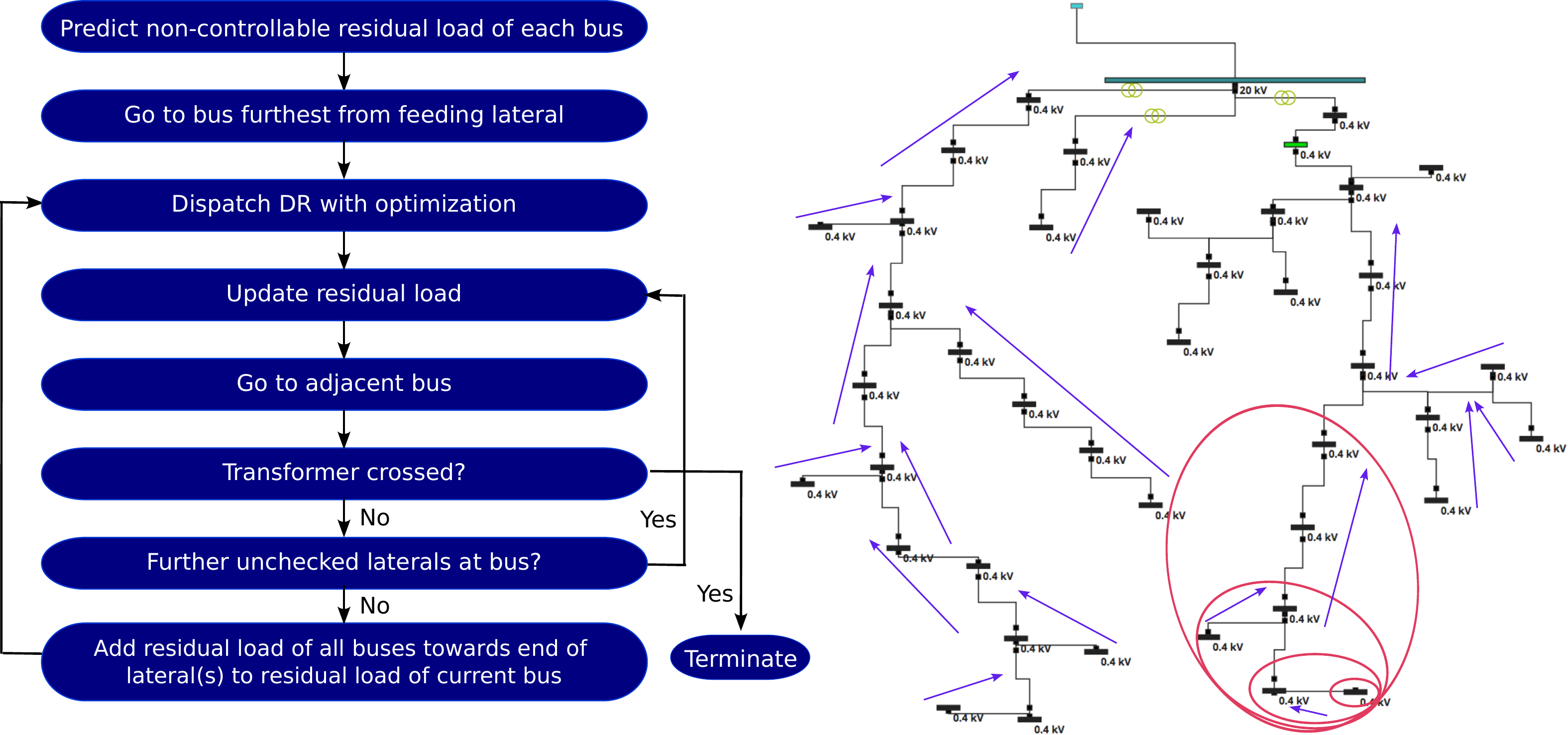} 
    \caption{Flow chart for heuristic-line-algorithm.}
    \label{fig:heuristic_line}
\end{figure}

\begin{figure}[!ht]
    \centering
    \includegraphics[height = 7cm]{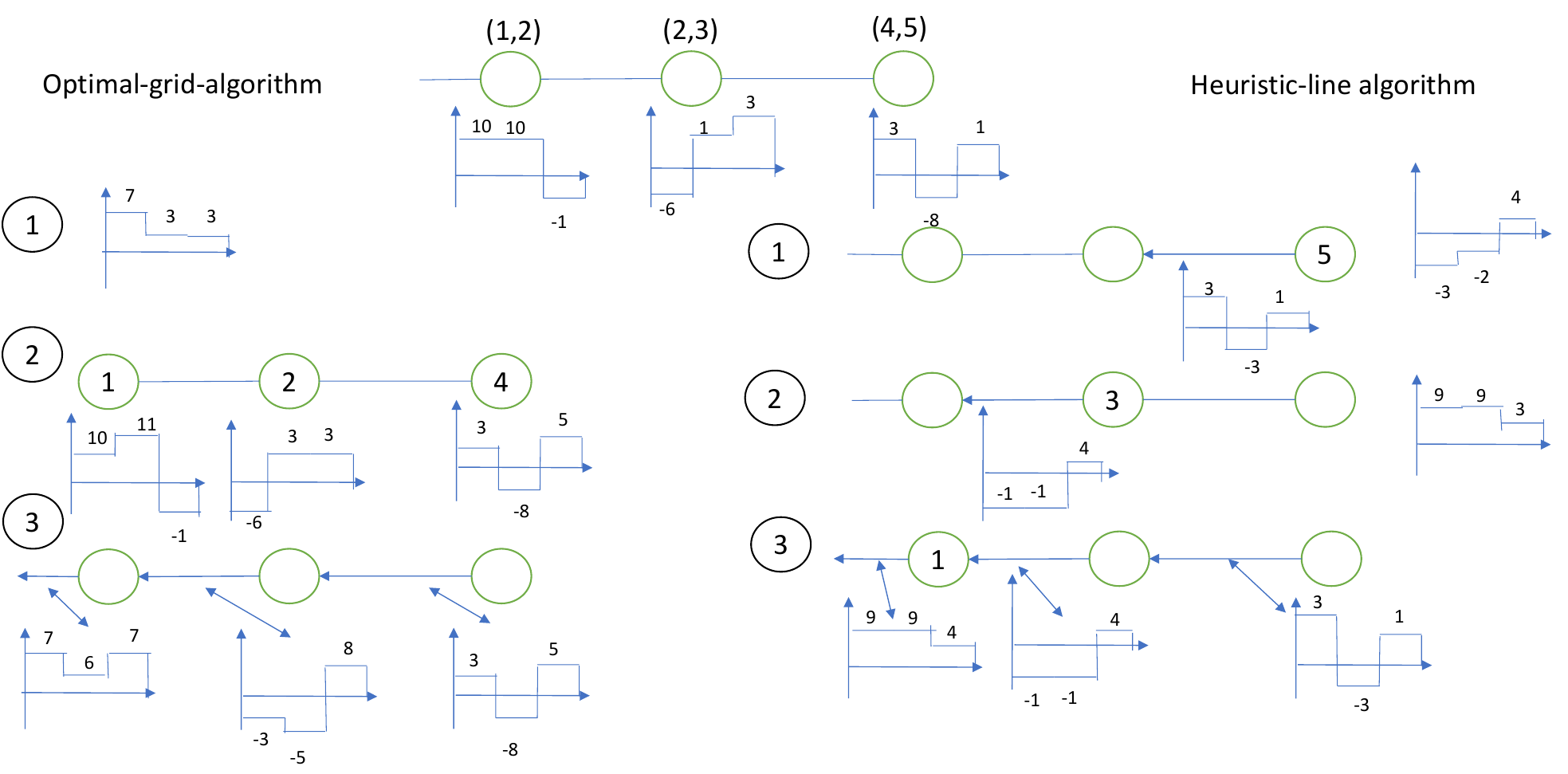} 
    \caption{Comparison of a few iterations of heuristic-line-algorithm and optimal-grid-algorithm.}
    \label{fig:algorithm_comparison}
\end{figure}

The heuristic-line-algorithm splits the original optimization problem into smaller sub-problems of lower complexity in the same fashion as the heuristic-grid-algorithm. In this case, the algorithm reduces the sum of peak power flows along each line in the grid. Again, the peak power flows are not optimally reduced. After predicting the non-controllable residual load at each bus, we start at the bus that is furthest away from the feeding transformer. All devices at this bus are dispatched so as to locally minimize the buses' local residual peak load. Then, we move to the adjacent bus towards the feeding transformer and add the already scheduled loads and non-controllable residual loads of the former bus to the current bus. Again, all controllable loads at the current bus are dispatched so as to locally minimize the peak of the updated residual load. In this backward-sweep, we move towards the feeding transformer in each grid and schedule devices at one bus after another. The optimization problem can be the same as for the heuristic-grid-algorithm. The main difference is, however, that devices are scheduled in a pre-defined sequence according to adjacency, using a more local residual load, that is, the aggregation of all "child-buses" towards end of lateral(s). Figure \ref{fig:heuristic_line} shows a flow chart of the algorithm. For a better understanding of the algorithm's behavior, we compare a few iterations with the optimal-grid-algorithm in Figure \ref{fig:algorithm_comparison}. 

The example in Figure \ref{fig:algorithm_comparison} considers three buses in radial structure. Each bus is assigned with its predicted non-controllable residual load over three time steps. The numbers in the brackets state how many units of power can or have to be scheduled from the DR resources on top of the residual profiles, so as to keep internal storage states within the consumer comfort constraints. The numbers in the black circles indicate the iteration step. The optimal-grid-algorithm will first aggregate the non-controllable residual load at each bus. Then, it schedules the required number of load units arbitrarily at each bus, but with the objective of reducing the peak power exchange on the outgoing line. Finally, we can see that the total peak power exchange with the rest of the grid is kept at 7, while consumer comfort is maintained. The heuristic-line-algorithm instead goes to the bus at the end of the branch and minimizes the residual peak load at the respective bus. By doing this, the initial negative load peak of -8 is turned into -3. Then, the algorithm moves on to the adjacent bus and adds the total residual load of the former bus to the non-controllable residual load of the current bus. The new negative and positive residual load peaks are -3 and 4. By adding two units of power in the first time step and one unit in the second time step, the negative peak is turned into -1. Compared to the optimal-grid-algorithm, we have achieved lower residual peaks in total while having higher peak power exchanges with the rest of the grid. 

\subsubsection{Critical-line-algorithm}

\begin{figure}[!ht]
    \centering
    \includegraphics[height = 8cm]{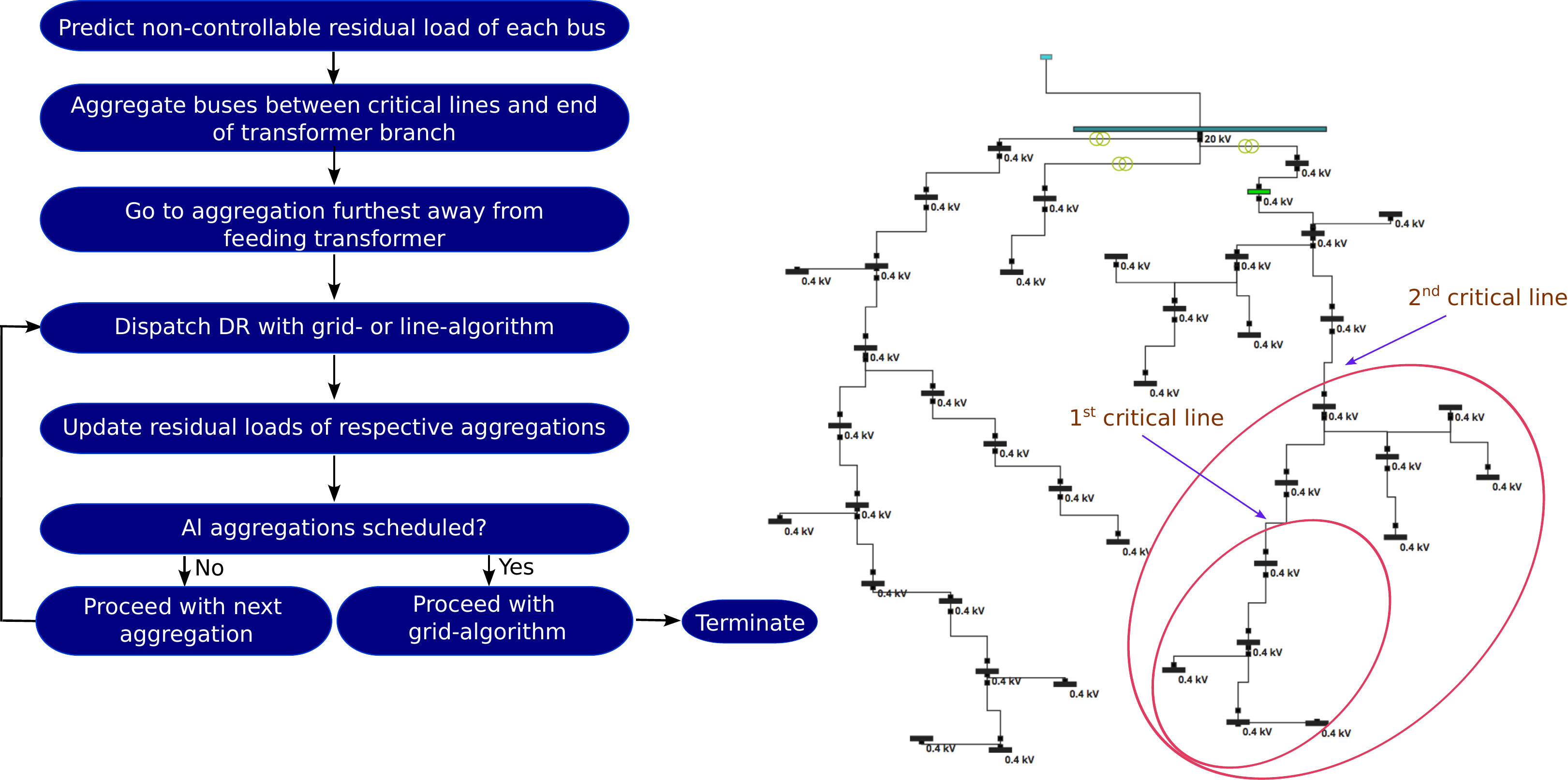} 
    \caption{Flow chart for critical-line-algorithm.}
    \label{fig:critical_line}
\end{figure}

The critical-line-algorithm combines the concepts of both heuristic algorithms we introduced above. An additional feature is that we now aggregate a number of buses between an as critical identified line and the end of branches and explicitly reduce the peak power exchange along the critical line with highest priority. Devices which are not in any aggregation are scheduled according to the residual load of their respective sub-grid. If no critical line is given and therefore no aggregation is made, the critical-line-algorithm behaves like the heuristic-grid-algorithm. Figure \ref{fig:critical_line} provides a flow chart for the algorithm.

\subsection{Empirical tests}

Our simulations have a temporal resolution of 15 minutes. All DR applications are generated with a high variety of parameters for input, output and loss behaviors, capacities, initial storage states and power ratings. All simulations start with the same random values for the distribution of prosumers (= producers + consumers), non-controllable residual loads and initial storage states. We further use a quadratic problem formulation, a sum-norm objective formulation and active and reactive power optimization for all simulations. We generate individual signals for each device with full control over the DR application, neglecting internal controllers of devices for the sake of simplicity. Consumer comfort constraints are relaxed with high weights so as to prioritize their maintenance. We use a simple 3.06 GHz Intel Core 2 Duo processor (iMac 2009) for our computation. We evaluate the performance of each algorithm based on computational time, grid losses, total active and reactive loads, residual active and reactive loads, maximum and sum of line and transformer loads, maximum and sum of voltage deviations at buses, and the sum over the magnitude of phase angle shifts.

In the randomized scenario, we simulate $T = 96$ time steps, corresponding to one day, with a total of 30 devices dispersed on the grid. Controllable loads represent 31\% of the total energy consumption in this period. The random dispersion leads to a maximum of 19 devices in the same transformer grid. Hence, our maximum computational time is here determined by the optimization of 19 devices at the same time. In the regular scenario, we simulate $T = 288$ time steps, corresponding to three consecutive days, with a total of 150 devices dispersed on the grid (neglecting air conditioners). Controllable loads this time represent 51\% of the total energy consumption in this period. The complexity of our DR control algorithms increases with the number of DR applications that we control and the number of time steps for which we optimize control signals into the future.

\section{Results}

\begin{figure}[!ht]
  \centering
    \includegraphics[height=210mm]{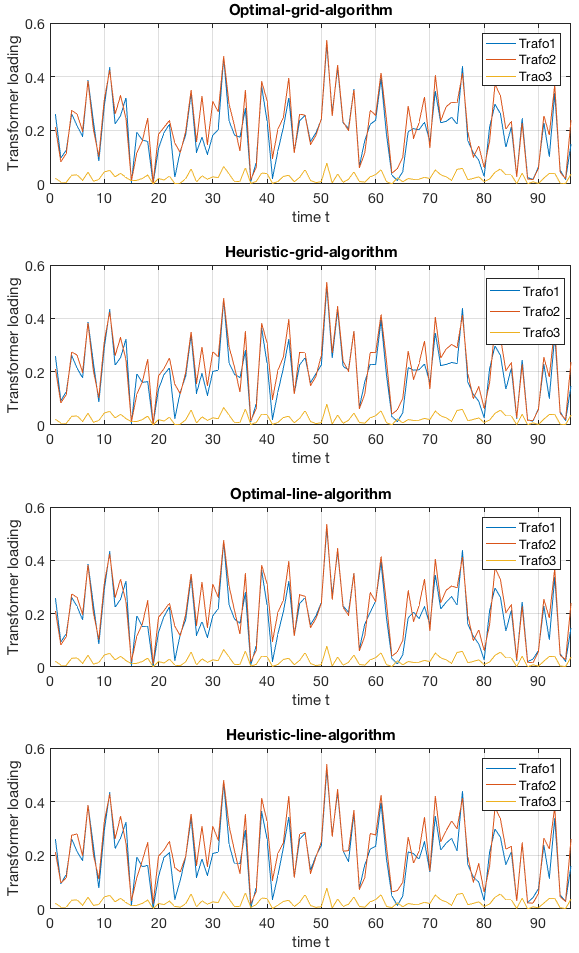}
    \caption{Transformer loading results for randomized scenario. 1 = 100\%}
    \label{fig:trafo_randomized}
\end{figure}

\begin{figure}[!ht]
  \centering
    \includegraphics[height=180mm]{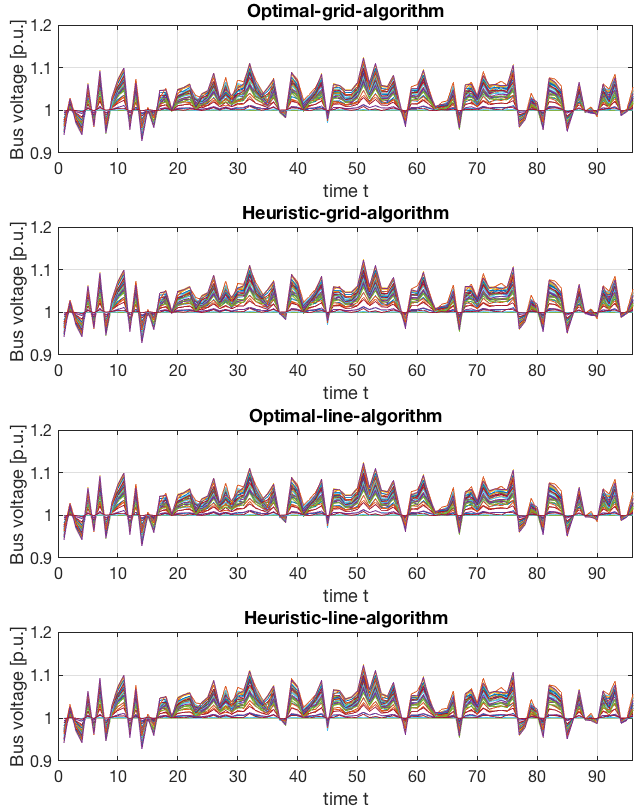}
    \caption{Bus voltage deviation results for randomized scenario. Each graph represents the voltage profile of one bus.}
    \label{fig:voltage_randomized}
\end{figure}

\begin{figure}[!ht]
  \centering
    \includegraphics[height=180mm]{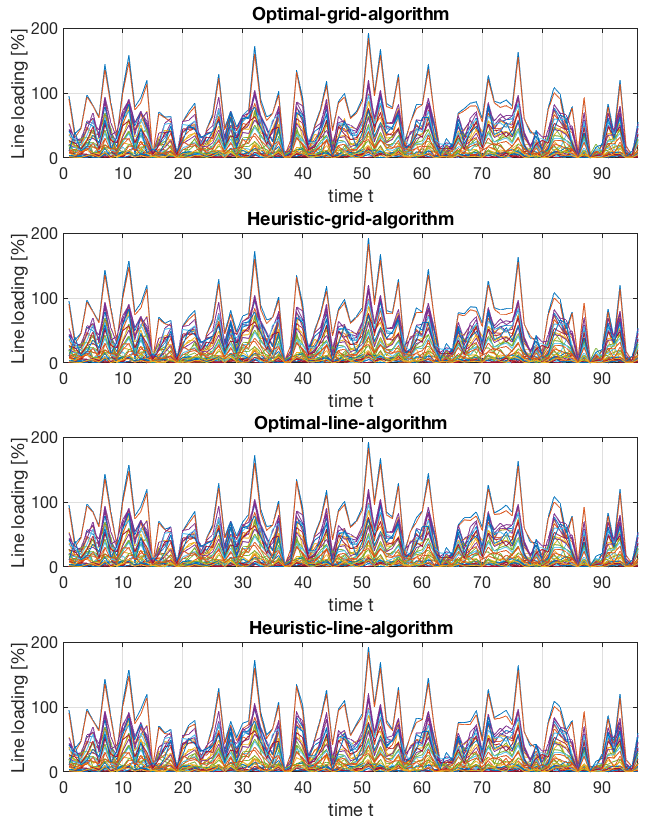}
    \caption{Line loading results for randomized scenario. Each graph represents the loading profile of one line.}
    \label{fig:line_randomized}
\end{figure}

\begin{figure}[!ht]
  \centering
    \includegraphics[height=180mm]{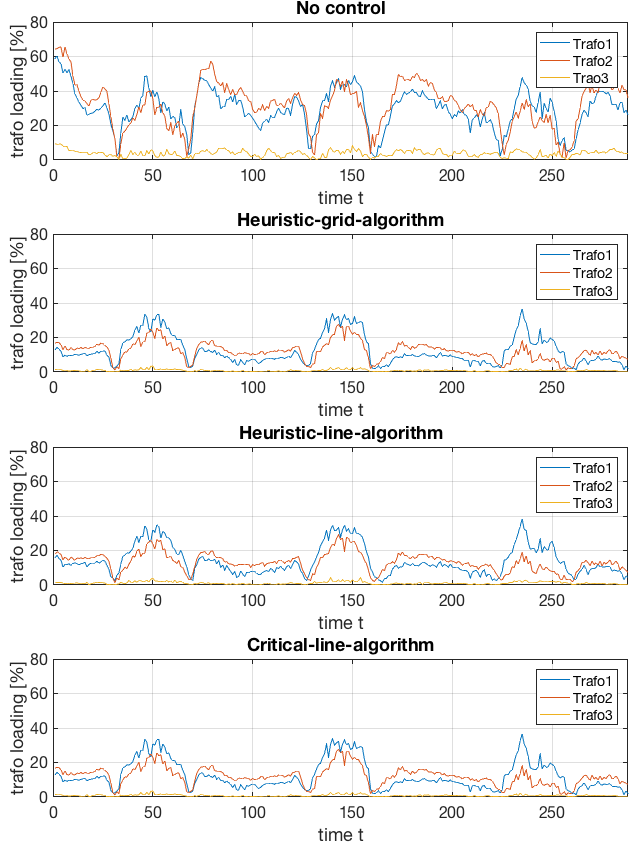}
    \caption{Transformer loading results for regular scenario.}
    \label{fig:trafo_regular}
\end{figure}

\begin{figure}[!ht]
  \centering
    \includegraphics[height=180mm]{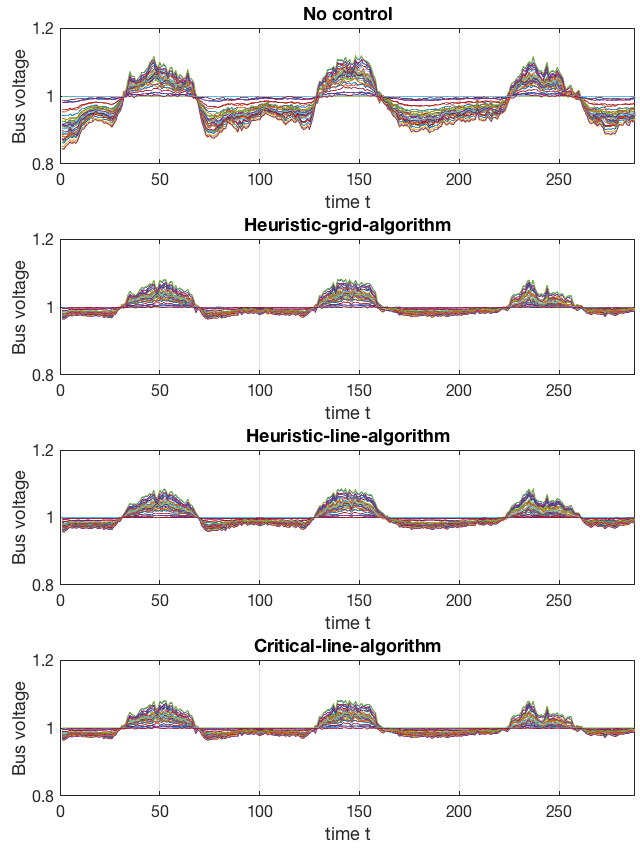}
    \caption{Bus voltage deviation results for regular scenario. Each graph represents the voltage profile of one bus.}
    \label{fig:voltage_regular}
\end{figure}

\begin{figure}[!ht]
  \centering
    \includegraphics[height=180mm]{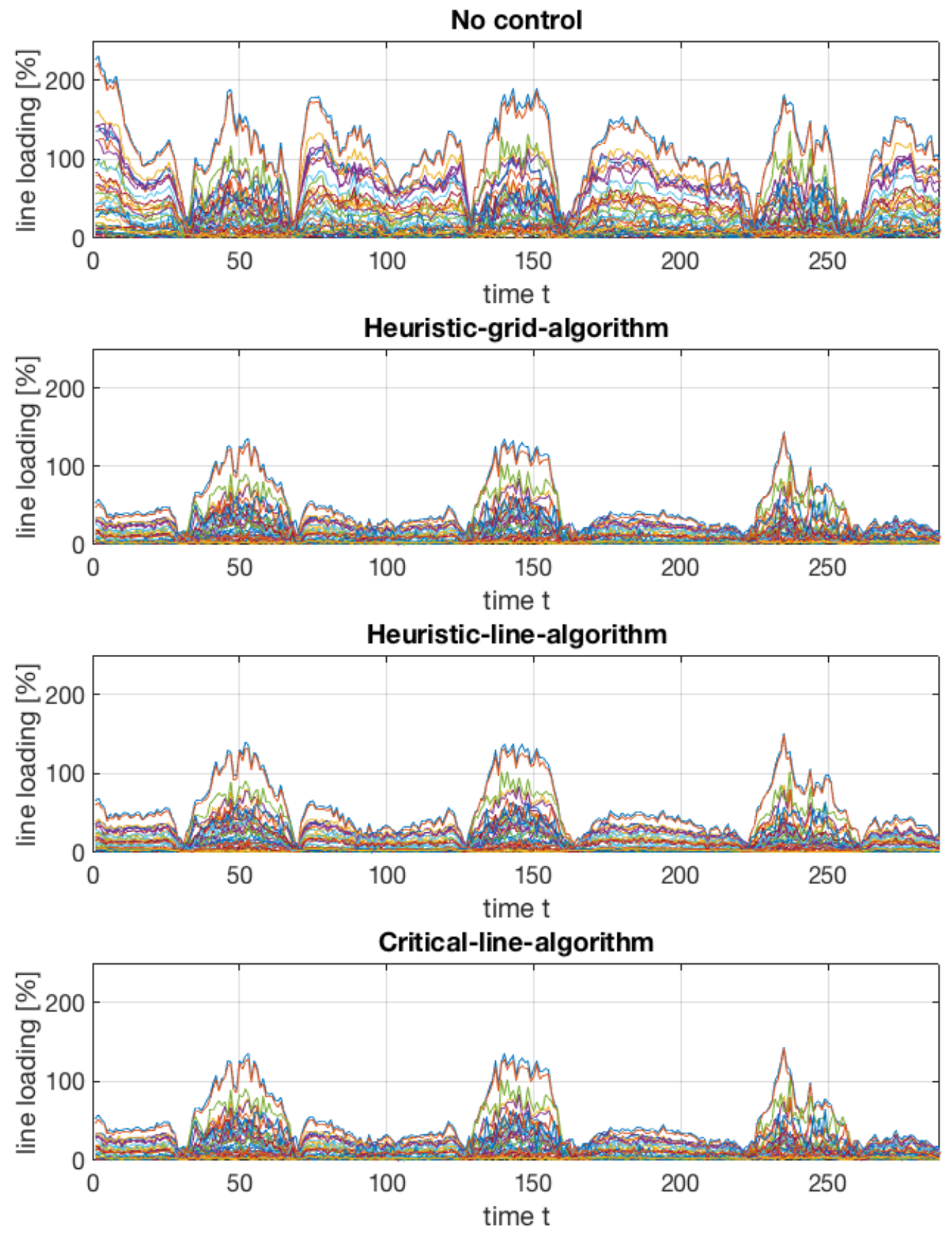}
    \caption{Line loading results for regular scenario. Each graph represents the loading profile of one line.}
    \label{fig:line_regular}
\end{figure}

Table \ref{tab:randomized_results} provides the numerical results of the empirical experiments for the randomized scenario. We compare the performance of the four main DR control algorithms against each other. Figures \ref{fig:trafo_randomized} - \ref{fig:line_randomized} further visualize resulting levels of transformer loading, bus voltage deviations and line loading. Table \ref{tab:regular_results} provides the numerical results of our empirical experiments for the regular scenario. Here, we only evaluate computationally traceable algorithms, that is, the heuristic algorithms, against a \emph{no control} scenario. Furthermore, we also test the performance of the critical-line-algorithm here. Figures \ref{fig:trafo_regular} - \ref{fig:line_regular} visualize the corresponding transformer loads, bus voltage deviations and line loads.

\subsection{Randomized scenario results}

The heuristic-grid/line-algorithms deliver results in about 39 seconds. In contrast, the optimal-grid/line-algorithms demand longer computational times of 10 and 16 minutes. The performance of all algorithms is similar. Both the grid- and line-algorithms reduce the peak power exchange with the higher level grid, that is, the transformer loading peaks, in a similar fashion. The total residual energy exchanges of the grids are also reduced to similar values. The heuristic-line-algorithm achieves less reduction in voltage deviations, phase angle shifts, transformer loading and grid losses than the other algorithms. Differences in the decimal places are likely caused by the optimizer's choice to cross relaxed consumer comfort constraints slightly differently whenever necessary for finding a solution. 

Although the optimal/heuristic-grid-algorithms are designed with the objective of reducing the peak residual power exchange with the higher voltage level, they seems to also reduce the sum of peak power flows along all lines in the same fashion as the optimal-line-algorithm. We conclude that reducing the total peak power exchange of the grid also reduces the sum of peak power flows along all lines for our examined radial grid topology. Additionally, scheduling one device at a time by solving a sequence of small optimization problems through the heuristic-grid-algorithm leads to the same system improvements at significantly lower computational time than scheduling all devices at the same time by solving one large optimization problem through the optimal-grid-algorithm. We conclude that the heuristic-grid-algorithm therefore outperforms the optimal-grid/line-algorithm. Furthermore, as the heuristic-grid-algorithm also leads to higher system improvements than the heuristic-line-algorithm, without requiring knowledge about grid topology, we conclude that the heuristic-grid-algorithm represents the best performing DR control algorithm we examine here.

\subsection{Regular scenario results}

The total energy consumption is reduced by about 20\% compared to the no control scenario, while consumer comfort constraints are strictly maintained. As a result of both lower total consumption and residual peak shaving, the active grid losses are reduced to around 1/4, maximum voltage deviation by about 3\%, maximum line loading by about 80\% and maximum transformer loading by about 30\%. The critical-line-algorithm performs similar to the heuristic-grid-algorithm on the entire grid, while requiring about 1.3 times the computational time. The maximum loading of the critical line is only slightly decreased. 

The critical-line-algorithm is designed to reduce peak loading along critical lines with higher priority. However, the performance differs only slightly from the other proposed algorithms, while again knowledge about grid topology is required and computational time increases by about 1.3 times. We conclude, that the proposed critical-line-algorithm does not bring useful improvements compared to the heuristic-grid-algorithm, even for prioritized lines.

\begin{center}
    \begin{table} [!ht]
        \caption{Simulation results for randomized scenario}
        \begin{tabular}{| l| l| l| l| l|}
        \hline
             & Optimal-grid & Heuristic-grid & Optimal-line & Heuristic-line\\
        \hline
             \multicolumn{5}{|c|}{General} \\
        \hline
             Computational time & 10 min  & 39 s & 16 min & 38 s\\
             Total load  MWh/MVArh &  9.9480  &  9.9522 & 9.9550  & 9.7919\\ 
             Residual load |MWh/MVArh| & 7.4263 & 7.4117 & 7.4201 & 7.6167 \\ 
             Active grid losses (sum) &  0.7461 MW & 0.7438 MW & 0.7446 MW & 0.7658 MW \\
        \hline
            \multicolumn{5}{|c|}{Buses} \\
        \hline
             Voltage deviation (|sum|) & 273.36\% &  272.83\% &  272.84\% &  280.34\% \\
             Voltage deviation (|max|) &  12.34\% &  12.34\% & 12.34\% & 12.34\% \\
             Phase angle shift (|sum|) & 39.7231$^\circ$ & 39.6455$^\circ$ & 39.6376$^\circ$ & 40.2030$^\circ$\\
        \hline
             \multicolumn{5}{|c|}{Lines} \\
        \hline 
             Loading (sum) &  2'082.13\% & 2'079.97\% &  2'076.47\% &  2'082.26\% \\
             Loading (max) & 192.16\% &  192.03\% & 192.04 \% & 192.20\%\\
        \hline 
             \multicolumn{5}{|c|}{Transformers} \\
        \hline 
             Loading (sum) &  1'395.01\% &  1'392.53\% &  1'394.47\% &  1'431.48\% \\
             Loading (max) & 53.68\% &  53.67\% &  53.67\% &  53.80\% \\
        \hline
        \end{tabular}
    \label{tab:randomized_results}
    \end{table}
\end{center}

\begin{center} 
    \begin{table} [!ht]
        \caption{Simulation results for all three algorithms vs. no control}
        \begin{tabular}{| l| l| l| l| l|}
        \hline
             & No Control & Heuristic-grid & Heuristic-line & Critical-line\\
        \hline
             \multicolumn{5}{|c|}{General} \\
        \hline
             Computational time & - & 17 min & 17 min & 23 min\\
             Total load  MWh/MVArh & 54.8952  & 43.1146 & 43.1520  & 43.0763\\ 
             Residual load |MWh/MVArh| & 32.1826 & 12.0699 & 13.9260  & 12.3465\\ 
             Active grid losses (sum) & 4.4011 MW & 0.8757 MW & 1.0487 MW & 0.8969 MW \\
        \hline
            \multicolumn{5}{|c|}{Buses} \\
        \hline
             Voltage deviation (|sum|) & 1'258.36 \% & 478.58 \% & 558.19 \% & 488.74\% \\
             Voltage deviation (|max|) &  11.8\% & 8.21\% &  8.61\% &  8.21\% \\
             Phase angle shift (|sum|) & 150.0995$^\circ$ & 90.5639$^\circ$ & 95.6108$^\circ$ & 90.2589$^\circ$\\
        \hline
             \multicolumn{5}{|c|}{Lines} \\
        \hline 
             Loading (sum) &  9'195.46\% &  4'080.25\% & 4'481.27\% &  4'121.45\% \\
             Loading (max) &  231.04\% &  143.83\% &  150.81\% & 143.83\%\\
        \hline 
             \multicolumn{5}{|c|}{Transformers} \\
        \hline 
             Loading (sum) &  6'295.87\% & 2'396.42\% & 2'741.99\% &  2'444.44\% \\
             Loading (max) &  65.55\% &  36.60\% &  38.35\% &  36.60\% \\
        \hline
             \multicolumn{5}{|c|}{Critical line 218874} \\
        \hline 
             Loading (sum) &  6'769.55\%& 1'576.15\% &  2'121.04\%& 1'669.18\%\\
             Loading (max) &  58.49\%&  18.69\%&  19.75\% & 18.69\%\\
        \hline
        \end{tabular}
    \label{tab:regular_results}
    \end{table}
    
\end{center}

\clearpage
\section{Discussion}

We investigate control objectives that provide an effective way for utilizing a large number of DR applications for the dispatch of active distribution networks, without violating consumer comfort or adversely affecting grid operation in the transmission network. We find that reducing the Euclidean-Norm of residual load at each transformer sub-station is an effective objective for utilizing DR for coordinating bi-directional power flows and increasing voltage stability in active distribution networks, while providing additional DR potentials for the dispatch of higher voltage levels. We show that this objective alleviates the need for information on grid topology and can be pursued without complex power flow calculations in real-time. Additionally, we observe that this objective reduces the sum of peak power flows along all lines in an active distribution network, and therefore results in lower system losses.

Our results provide an answer to a long-lasting research problem, that is, the problem of translating complex power system calculations into simplified control objectives. Our findings are consistent with observations in previous studies, where DR for example leads to 30\% lower system losses \cite{Gupta2022}. 

Further research in this direction, however, is not advisable as machine learning based approaches provide increasingly efficient tools that outperform traditional control algorithms as the ones we propose \cite{Donti2022, Aryandoust2023}. Instead, further research can contribute to the collection of machine learning tasks and datasets for frequency control, optimal power flow and phase imbalance control, as well as the simultaneous dispatch of inverters, capacitors and renewable generation curtailment. This would further facilitate the design of artificial intelligence and machine learning models in this domain \cite{Aryandoust2023_datasets}.

\section*{Appendix}

In this appendix, we show how a reformulation of losses along lines, in-phase transformers and phase-shifting transformers can be derived from basic power flow equations.

\textit{Lines:}

\begin{equation}
    \begin{split}
    S_{km} &= E_k I_{km}* \\
        & = U_k e^{j\theta_k}(y_{km}^* (U_k e^{-j\theta_k} - U_m e^{-j\theta_m}) + y_{km}^{sh*} U_k e^{-j\theta_k} \\
        & = U_k^2 y_{km}^* - U_k U_m y_{km}^* e^{j\theta_{km}} + U_k^2 y_{km}^{sh*} \\
        & = U_k^2 (g_{km} - j b_{km}) - U_k U_m e^{j\theta_{km}} (g_{km} - j b_{km}) - j U_k^2 b_{km}^{sh} \\
        & = U_k^2 g_{km} - j U_k^2 b_{km} - U_k U_m g_{km} e^{j\theta_{km}} + j U_k U_m b_{km} e^{j\theta_{km}} - j U_k^2 b_{km}^{sh} \\
        & = U_k^2 g_{km} - j U_k^2 b_{km} - U_k U_m g_{km} cos\theta_{km} - j U_k U_m g_{km} sin\theta_{km} \\
        & + j U_k U_m b_{km} cos\theta_{km} - U_k U_m b_{km} sin\theta_{km} - j U_k^2 b_{km}^{sh} \\
        & = U_k^2 g_{km} - U_k U_m g_{km} cos\theta_{km} - U_k U_m b_{km} sin\theta_{km} \\
        & + j [ - U_k^2 b_{km} + U_k U_m b_{km} cos\theta_{km} - U_k U_m g_{km} sin\theta_{km} - U_k^2 b_{km}^{sh} ] \\
    P_{km} & = U_k^2 g_{km} - U_k U_m g_{km} cos\theta_{km} - U_k U_m b_{km} sin\theta_{km} \\
    Q_{km} & = - U_k^2 (b_{km} + b_{km}^{sh}) + U_k U_m b_{km} cos\theta_{km} - U_k U_m g_{km} sin\theta_{km} \\
    \end{split}
\end{equation}

$S_{mk}$ is straightforward: replace $\theta_{km}$ with $\theta_{mk}$ and $U_k$ with $U_m$:

\begin{equation}
    \begin{split}
    P_{mk} & = U_m^2 g_{km} -U_k U_m g_{km} cos\theta_{mk}  - U_k U_m b_{km} sin\theta_{mk} \\
    Q_{mk} & = - U_m^2 (b_{km} + b_{km}^{sh}) + U_k U_m b_{km} cos\theta_{mk} - U_k U_m g_{km} sin\theta_{mk} \\ 
    \end{split}
\end{equation}

The losses along a line can then be calculated and reformulated in dependence on the magnitude of voltage drop $|E_k - E_m|$:

\begin{equation}
    \begin{split}
    P_{loss}^{line} & = P_{km} + P_{mk} \\
    & = g_{km}   [ U_k^2 + U_m^2 - U_k U_m   (cos\theta_{km} +cos\theta_{mk})] \\ 
    & - U_k  U_m  b_{km}   (sin\theta_{km} + sin\theta_{mk}) \\
    & = g_{km}   (U_k^2 + U_m^2 - 2  U_k  U_m cos\theta_{km}) \\
    & = g_{km}   |E_k - E_m|^2 \\
    Q_{loss}^{line} & = Q_{km} + Q_{mk} \\
    & = - (b_{km}^{sh} + b_{km})   (U_m^2 + U_k^2) + U_k U_m b_{km}   (cos\theta_{km} + cos\theta_{mk}) \\
    & - U_k U_m g_{km}   (sin\theta_{km} + sin\theta_{mk}) \\
    & = - b_{km}^{sh}   (U_m^2 + U_k^2) - b_{km}   (U_k^2 + U_m^2 - 2 U_k U_m cos\theta_{km}) \\
    & = - b_{km}^{sh}   (U_m^2 + U_k^2) - b_{km}   |E_k - E_m|^2
    \end{split}
\end{equation}

\textit{In-phase transformer:}

\begin{equation}
    \begin{split}
    S_{km} & = E_k I_{km}^* \\
        & = U_k e^{j\theta_k} (a_{km}^2 y_{km}^* U_k e^{-j \theta_k} - a_{km} y_{km}^* U_m e^{-j \theta_m}) \\
        & = a_{km}^2 U_k^2 y_{km}^* - a_{km} U_k U_m y_{km}^* e^{j\theta_{km}} \\
        & = a_{km}^2 U_k^2 (g_{km} - j b_{km}) - a_{km} U_k U_m e^{j\theta_{km}} (g_{km} - j b_{km}) \\
        & = U_k^2 a_{km}^2 g_{km} - j b_{km} U_k^2 a_{km}^2 - a_{km} U_k U_m g_{km} e^{j\theta_{km}} + j U_k U_m a_{km} b_{km} e^{j\theta_{km}} \\
        & =  U_k^2 a_{km}^2 g_{km} - j b_{km} U_k^2 a_{km}^2 - a_{km} U_k U_m g_{km} cos\theta_{km} - j a_{km} U_k U_m g_{km} sin\theta_{km} \\
        & + j U_k U_m a_{km} b_{km} cos\theta_{km} - U_k U_m a_{km} b_{km} sin\theta_{km} \\
        & = (U_k a_{km})^2 g_{km} - a_{km} U_k U_m g_{km} cos\theta_{km} - a_{km} U_k U_m b_{km} sin\theta_{km} \\
        & + j [ -(U_k a_{km})^2 b_{km} - a_{km} U_k U_m g_{km} sin\theta_{km} + U_k U_m a_{km} b_{km} cos\theta_{km} ] \\
    P_{km} & =  (a_{km} U_k)^2 g_{km} - a_{km} U_k U_m g_{km} cos\theta_{km} - a_{km} U_k U_m b_{km} sin\theta_{km} \\
    Q_{km} & = - (a_{km} U_k )^2 b_{km} + a_{km} U_k U_m  b_{km} cos\theta_{km} - a_{km} U_k U_m g_{km} sin\theta_{km} 
    \end{split} 
\end{equation}

$S_{mk}$ is straightforward: replace $\theta_{km}$ with $\theta_{mk}$ and $a_{km}U_k$ with $U_m$:

\begin{equation}
    \begin{split}
    P_{mk} & = U_m^2 g_{km} -a_{km}  U_k U_m g_{km} cos\theta_{mk} - a_{km} U_k U_m b_{km} sin\theta_{mk} \\ 
    Q_{mk} & = - U_m^2 b_{km} + a_{km}  U_k U_m b_{km} cos\theta_{mk} - a_{km} U_k U_m g_{km} sin\theta_{mk} \\ 
    \end{split}
\end{equation}

The losses along an in-phase transformer can then be calculated and reformulated in dependence on the magnitude of voltage drop $|E_k - E_m|$:

\begin{equation}
    \begin{split}
    P_{loss}^{in-ph} & = P_{km} + P_{mk} \\
        & = g_{km} [a_{km}^2 U_k^2 + U_m^2 - 2 a_{km} U_k U_m cos\theta_{km}] - a_{km} b_{km} U_k U_m (sin\theta_{km} + sin\theta_{mk}) \\
        & = g_{km} |a_{km} E_k - E_m|^2 \\
    Q_{loss}^{in-ph} & = Q_{km} + Q_{mk} \\
        & = -b_{km} (a_{km}^2 U_k^2 + U_m^2) + 2 U_k U_m a_{km} b_{km} cos\theta_{km} - U_k U_m a_{km} g_{km} (sin\theta_{km} + sin\theta_{mk}) \\
        & = - b_{km} (a_{km}^2 U_k^2 + U_m^2 - 2 U_k U_m a_{km} cos\theta_{km}) \\
        & = b_{km} |a_{km} E_k - E_m|^2 
    \end{split} 
\end{equation}

\textit{Phase-shifting transformer:} 

\begin{equation}
    \begin{split}
    S_{km} &= E_k I_{km}^* \\
        & = U_k e^{j\theta_k} (a_{km}^2 y_{km}^* U_k e^{-j \theta_k} - t_{km} y_{km}^* U_m e^{-j \theta_m} \\
        & = U_k^2 a_{km}^2 y_{km}^* - U_k U_m t_{km} y_{km}^* e^{j \theta_{km}} \\
        & = U_k^2 a_{km}^2 (g_{km} - j b_{km}) - U_k U_m t_{km} e^{j \theta_{km}} (g_{km} - j b_{km} ) \\
        & = U_k^2 a_{km}^2 g_{km} - j U_k^2 a_{km}^2 b_{km} - U_k U_m a_{km} g_{km} e^{j(\theta_{km} + \phi_{km})} + j U_k U_m a_{km} b_{km} e^{j(\theta_{km} + \phi_{km})} \\
        & = U_k^2 a_{km}^2 g_{km} - j U_k^2 a_{km}^2 b_{km} - U_k U_m a_{km} g_{km} cos(\theta_{km} + \phi_{km}) - j U_k U_m a_{km} g_{km} sin(\theta_{km} + \phi_{km}) \\
        & + j U_k U_m a_{km} b_{km} cos(\theta_{km} + \phi_{km}) - U_k U_m a_{km} b_{km} sin(\theta_{km} + \phi_{km}) \\
    P_{km} & = U_k^2 a_{km}^2 g_{km} - U_k U_m a_{km} g_{km} cos(\theta_{km} + \phi_{km}) - U_k U_m a_{km} b_{km} sin(\theta_{km} + \phi_{km}) \\
    Q_{km} &= - U_k^2 a_{km}^2 b_{km} + U_k U_m a_{km} b_{km} cos(\theta_{km} + \phi_{km})  - U_k U_m a_{km} g_{km} sind(\theta_{km} + \phi_{km}) \\
    \end{split}
\end{equation}

$S_{mk}$ is straightforward: replace $\theta_{km}$ with $\theta_{mk}$, $\phi_{km}$ with $-\phi_{km}$  and $a_{km}U_k$ with $U_m$:

\begin{equation}
    \begin{split}
    P_{mk} & = U_m^2 g_{km} -a_{km}  U_k U_m g_{km} cos(\theta_{mk} -\phi_{km}) - a_{km} U_k U_m b_{km} sin(\theta_{mk} - \phi_{km}) \\ 
    Q_{mk} & = - U_m^2 b_{km} + a_{km}  U_k U_m b_{km} cos(\theta_{mk} - \phi_{km}) - a_{km} U_k U_m g_{km} sin(\theta_{mk} - \phi_{km}) \\ 
    \end{split}
\end{equation}

The losses along a phase-shifting transformer can then be calculated and reformulated in dependence on the magnitude of voltage drop $|E_k - E_m|$:

\begin{equation}
    \begin{split}
    P_{loss}^{ph-shift} & = P_{km} + P_{mk} \\
    & = g_{km} (U_k^2 a_{km}^2 + U_m^2 - U_k U_m a_{km} (cos(\theta_{km} + \phi_{km}) + cos(\theta_{mk} - \phi_{km})) \\
    & - U_k U_m a_{km} b_{km} (sin(\theta_{km} + \phi_{km}) + sin(\theta_{km} + \phi_{km}) + sin(\theta_{km} - \phi_{km})) \\
    & = g_{km} (U_k^2 a_{km}^2 + U_m^2 - 2 U_k U_m a_{km} cos(\theta_{km} + \phi_{km})) \\
    & = g_{km} |t_{km} E_k - E_m|^2 \\
    Q_{loss}^{ph-shift} & = Q_{km} + Q_{mk} \\
    & = - b_{km} (U_k^2 a_{km}^2 + U_m^2) + 2 U_k U_m a_{km} b_{km} cos(\theta_{km} + \phi_{km}) \\
    & - U_k U_m a_{km} g_{km} (sin(\theta_{km} + \phi_{km}) + sin(\theta_{mk} - \phi_{km})) \\
    & = - b_{km} ( U_k^2 a_{km}^2 + U_m^2 - 2 U_k U_m a_{km} b_{km} cos(\theta_{km} + \phi_{km})) \\
    & = - b_{km} |t_{km} E_k - E_m|^2 
    \end{split}
\end{equation}

\end{document}